\documentclass[10pt,fleqn,a4paper]{article}

\usepackage{graphicx}
\usepackage{amsfonts}
\usepackage[mathscr]{euscript}
\usepackage{here}
\usepackage[numbers]{natbib}
\usepackage{notoccite}

\newcommand{\bb}[1]{\mbox{\boldmath $#1$\unboldmath}}

\newcommand{\dd}{\displaystyle}
\newcommand{\vp}{\vspace{4pt}}

\newcommand{\tr}{\makebox{tr}}

\renewcommand\appendix{\par
  \setcounter{section}{0}
  \setcounter{subsection}{0}
  \setcounter{figure}{0}
  \setcounter{table}{0}
  \renewcommand\thesection{Appendix \Alph{section}}
  \renewcommand\thefigure{\Alph{section}\arabic{figure}}
  \renewcommand\thetable{\Alph{section}\arabic{table}}
}

\title{On the structure of metal plasticity constitutive equations and the physical origin of the plastic spin}

\author{Stefan C. Soare\,\,\thanks{Technical Univ. of Cluj-Napoca, \, stef$\_$soare@yahoo.com}}

\date{\small{27 June 2013}}

\begin{document}
\maketitle

\begin{abstract}
The macroscopic elastic-plastic response of a metal polycrystal is
analyzed here by homogenizing the response of a representative
volume element. With this purpose, the theory developed in
Hill(1967)(The essential structure of constitutive laws for metal
composites and polycrystals. J. Mech. Phys. Solids, 15, 79-95) is
extended so that the structural changes caused by plastic
deformation are taken into account. It is found that the type of
evolution assigned to the slip systems of a constituent crystal
influences significantly the structure of the constitutive system
characterizing the overall response. In particular, the classical
normality structure is obtained only if slip systems are assumed
to convect with the crystal lattice. Any other type of slip system
evolution induces a deviation of the macro-rate of plastic
deformation from the exterior normal to the yield surface. This
provides a physical explanation and also a rigorous definition for
the much debated concept of plastic spin.
\end{abstract}


\section{Introduction}

\noindent In the macroscopic (rate-independent) theory of metal
plasticity the yield surface characterizes the plastic properties
of a material: it bounds the elastic domain, the set of stress
states for which the material responds elastically, and it
determines via the associated flow rule the direction of the rate
of plastic deformation. A correct and accurate modeling of the
yield surface is therefore of primary importance, particularly
when plastic properties are anisotropic. In this case, the yield
surface is necessarily described with respect to some material
frame, or is expressed as an invariant of the stress state and
of a set of structural tensors, e.g., \cite{Liu},
\cite{Boehler}.

To illustrate the issue, consider the case of a rolled thin sheet
for which one can discern, in some initial state, three orthogonal
axes of material symmetry, the rolling, transverse and normal
directions. The material frame may be aligned, initially, along
these axes and the yield surface may be described analytically
with respect to this frame. Supposing the sheet is further subject
to some deformation process, the question is: how does the
material frame evolve during this deformation ? One plausible
choice would be to let the material frame rotate at the spin of
the motion, or, alternatively, let it rotate with the rotational
part of the deformation gradient, \cite{Tugcu}. This is certainly
correct when texture evolution is neglected, the hardening being
isotropic, and hence the shape of the yield surface is not
affected by deformation. In general, however, plastic deformation
is associated with texture evolution and hence the shape of the
yield surface does evolve with deformation and even initial
symmetries may be lost. The above question can now be
reformulated: in the general case, when texture evolution is taken
into account, does the material frame rotate with the material
(say, at the spin of the motion) or does it feature an additional
rotation with respect to the material ? It will be shown here that
both point of views are equally valid and reflect, in fact,
different characterizations of the yield surface.

The second possibility mentioned above deserves further comments.
As indicated by the experiments reported by Kim and Yin\cite{KimYin},
it may happen that the plastic symmetry of the material is not completely destroyed when particular paths of deformation are followed.
In these cases, within a certain degree of approximation,
it can be considered that the symmetry axes, and hence the material frame, rotate
rigidly with respect to the material, while the shape of the yield surface varies with respect to the material frame, \cite{Bunge97}.
The rate at which this relative rotation takes place has been called plastic spin.
An illustration in the context of Kim and Yin\cite{KimYin} experiments can be found in \cite{Dafalias2000} (where yield surface distortion is neglected).

Although there is quite a significant body of literature
concerning the plastic spin, no rigorous justification of the
concept has been proposed yet.
In general, the arguments are heuristic and involve
a significant degree of approximation. As representative works in this direction let us mention:
Mandel\cite{Mandel82}, where the plastic spin is defined as the direct average
of the lattice spin of the constituent crystals;
this "definition" seems to have
propagated in many works dedicated to the study of texture
evolution, e.g., \cite{Prantil93}, \cite{Bunge97};
Dafalias and Aifantis\cite{Dafalias90}, where an ab nihilo scale invariance argument is employed
to "deduce" the structure of the macroscopic model of plastic flow.
Further arguments for the plastic spin as macro-variable of the theory of
plasticity are purely phenomenological, the works of Dafalias
being representative for this approach, see \cite{Dafalias98} for
a relatively recent account. Also notable in the cited work, as
well as in that of Mandel\cite{Mandel82}, are the precautions taken by
the author regarding the additive decomposition of the rate of
deformation into elastic and plastic parts, about which objective
stress-rate is most appropriate, and the belief that the additive
decomposition can be avoided by employing an elastic-plastic multiplicative
decomposition (of the deformation gradient) together with an
explicitly hyper-elastic response. These, however, are sources for
further confusion since, given any multiplicative decomposition or
any stress-strain relationship, hyper-elastic or not, one can
always take its time derivative and thus obtain an additive
decomposition.


At the root of all of the above problems lies an incomplete
understanding of the relationship between the macro-rate of
plastic deformation and the yield surface, or, more fundamentally,
the lack of rigorous interpretations of the macro-rate of plastic
deformation and of the yield surface in terms of microstructure.
In contrast, phenomenological theories of crystal plasticity enjoy
an unequivocal definition of the rate of plastic deformation, the
Taylor\cite{Taylor38} representation, which leads to a precise
characterization of its relationship with the activation surface.
The first major attempt at linking the two scales, in an
elastic-plastic context, is that of Hill\cite{Hill67}.
However, the cited work neglects precisely the kind of aspects
that are of interest in the present context: the structural
changes caused by plastic deformation, or, texture evolution.
Here, the theory in \cite{Hill67} is reconsidered and extended so that structural changes are taken into account,
the goal remaining the same: to deduce the essential structure of the macroscopic elastic-plastic response of a polycrystal aggregate.

\subsection{Notations}

\noindent Second and fourth order tensors are also viewed as
linear operators and the symbol "$:$" is used to denote the
application of an operator to its argument. Thus, if $\bb{K}$ and
$\bb{e}$ are fourth and second order tensors, respectively, with
components $K_{ijkl}$ and $e_{kl}$ with respect to some
orthonormal basis, then $\bb{\sigma}=\bb{K}:\bb{e}$ defines the
second order tensor $\bb{\sigma}$ with components $\sigma_{ij} :=
K_{ijkl}e_{kl}$\,; the symbol "$:=$" emphasizes a definition. The
dot product, between tensors of the same kind, is always denoted
by a single dot, e.g. $\bb{\sigma}\cdot\bb{I} =
\sigma_{ij}\delta_{ij}=\tr(\bb{\sigma})$. The symbol $\otimes$
denotes the tensor product: $(\bb{a}\otimes\bb{b}):\bb{c} :=
(\bb{b}\cdot\bb{c})\bb{a}$. Finally, $\bb{A}^T$ denotes the
transposed operator of $\bb{A}$.

\section{Single crystal: basic equations}

\noindent Let $x$ denote the position vector of a particle of a
crystal, measured with respect to some spatial (fixed) reference
frame; with the velocity field $\bb{v}(x,t)$ at the moment $t$ we
associate the velocity gradient $\bb{l}(x,t) =
\partial\bb{v}/\partial x$. It is assumed that the
only mechanism of plastic deformation at crystal level is by
plastic shear along specific crystallographic directions,
Taylor\cite{Taylor38}. Then letting $\bb{m}^{\alpha}$ and
$\bb{n}^{\alpha}$ denote the slip direction and the normal to a
slip plane, the velocity gradient admits during elastic-plastic
deformation the representation
\begin{equation}
\bb{l} = \bb{l}^e + \widehat{\bb{l}}^p,\,\,\,\makebox{where}\,\,\,\,
\widehat{\bb{l}}^p = \sum_{\alpha\in\mathcal{A}}\dot{\gamma}^{\alpha}\bb{g}^{\alpha},\,\,\,\,
\bb{g}^{\alpha} = \bb{m}^{\alpha}\otimes\bb{n}^{\alpha}
\label{vgradient}
\end{equation}
where $\dot{\gamma}^{\alpha}$ denotes the rate of plastic shear on
slip system $\alpha\in\mathcal{A}\subset\mathcal{S}$; by
$\mathcal{A}$ is denoted the set of active slip systems at the
moment $t$ and by $\mathcal{S}$ the set of all slip systems at
particle $x$. By taking the symmetric and antisymmetric parts of
the above representation, the rate of deformation
$\bb{d}=(\bb{l}+\bb{l}^T)/2$ and the material spin
$\bb{w}=(\bb{l}-\bb{l}^T)/2$ acquire the decompositions:
\begin{equation}
\bb{d} = \bb{d}^e + \widehat{\bb{d}}^p,\,\,\,\, \widehat{\bb{d}}^p =\sum_{\alpha\in\mathcal{A}}\dot{\gamma}^{\alpha}\bb{a}^{\alpha},\,\,\,\,
\bb{a}^{\alpha} = \left(\bb{g}^{\alpha}+\bb{g}^{\alpha T}\right)/2
\label{ratedef}
\end{equation}
\begin{equation}
\bb{w} = \bb{w}^e + \bb{w}^p,\,\,\,\, \bb{w}^p =\sum_{\alpha\in\mathcal{A}}\dot{\gamma}^{\alpha}\bb{b}^{\alpha},\,\,\,\,
\bb{b}^{\alpha} = \left(\bb{g}^{\alpha}-\bb{g}^{\alpha T}\right)/2
\label{ratedef2}
\end{equation}

Letting $\bb{\tau}:=J\bb{\sigma}$ denote the Kirchhoff stress,
where $J$ is the determinant of the deformation gradient and $\bb{\sigma}$ is the Cauchy stress,
the stress-strain relationship at particle $x$ reads, \cite{HillRice72}, \cite{AsaroRice},
\cite{Asaro}:
\begin{equation}
\bb{\dot{\tau}}^{Je}=\bb{\dot{\tau}}+\bb{\tau}\bb{w}^e-\bb{w}^e\bb{\tau} = \bb{K}^J:\left(\bb{d}-\widehat{\bb{d}}^p\right)
\label{jaumanne}
\end{equation}
A superposed dot denotes the material derivative; $\bb{\dot{\tau}}^{Je}$ is the Jaumann rate associated with the spin $\bb{w}^e$
and $\bb{K}^J$ the corresponding tensor of instantaneous moduli.
Thus the above equation reproduces, with respect to the spatial frame,
the point of view of an observer rotating rigidly with the crystal lattice (lattice frame).
With $\bb{w}^e=\bb{w}-\bb{w}^p$, for an observer in a frame rotating with the crystal
(at spin $\bb{w}$) the stress-strain relationship reads:
\begin{equation}
\bb{\dot{\tau}}^{J}=\bb{\dot{\tau}}+\bb{\tau}\bb{w}-\bb{w}\bb{\tau} = \bb{K}^J:\left(\bb{d}-\bb{d}^p\right)
\label{jaumann}
\end{equation}
where the rate of plastic deformation associated with the material
frame is
\begin{equation}
\bb{d}^p = \widehat{\bb{d}}^p - \left(\bb{K}^J\right)^{-1}:\left(\bb{\tau}\bb{w}^p-\bb{w}^p\bb{\tau}\right)
\label{jaumanndp}
\end{equation}

More generally, to each (objective) stress rate there corresponds
a specific rate of plastic deformation. In other words, the rate
of plastic deformation, being a rate, depends on the frame of the
observer. However, once the constitutive relationships in
eqs.(\ref{vgradient}) and (\ref{jaumanne}) are given, the rate of
plastic deformation and the stress-strain relationship are
uniquely determined with respect to any frame. To formalize these
ideas, consider an objective stress rate $\bb{\dot{\tau}}^Z$
defined by
\begin{equation}
\bb{\dot{\tau}}^Z:=\bb{\dot{\tau}}^J-\bb{z}(\bb{\tau},\bb{d}),\,\,\,\, \makebox{with}\,\,\,\,
\bb{z}(\bb{\tau},\bb{d}) = \bb{Z}(\bb{\tau}):\bb{d}
\end{equation}
where $\bb{Z}$ is an isotropic fourth order tensor. In particular, $\bb{Z}$ is symmetric.
The above representation is general enough to include many of the known objective rates, \cite{GurtinAnand},
and also the rate of any stress measure work-conjugated with an arbitrary strain measure defined
with respect to some reference configuration, \cite{Hill68a}. Employing eq.(\ref{jaumann}), an observer
associated with the above rate writes
\begin{equation}
\bb{\dot{\tau}}^Z = \bb{K}^Z:\left(\bb{d}-\bb{d}^{pZ}\right)
\label{jaumannz}
\end{equation}
with $\bb{K}^Z$ and $\bb{d}^{pZ}$ defined by
\begin{equation}
\bb{K}^Z := \bb{K}^J-\bb{Z},\,\,\,\,\bb{d}^{pZ}:=\left(\bb{K}^Z\right)^{-1}\bb{K}^J:\bb{d}^{p}
\label{jaumannzdp}
\end{equation}
Note that if $\bb{K}^J$ is (major) symmetric, $\bb{K}^Z$ is also symmetric;
it will be assumed that $\bb{K}^Z$ is also invertible.
Employing eqs.(\ref{jaumanndp}) and (\ref{jaumannzdp}), the relationship between $\bb{d}^{pZ}$ and $\widehat{\bb{d}}^p$ reads
\begin{equation}
\bb{d}^{pZ} = \widehat{\bb{d}}^p +
\left(\bb{K}^Z\right)^{-1}:\left[\bb{z}(\bb{\tau},\widehat{\bb{d}}^p)+\bb{\tau}\bb{w}^{pT}+\bb{w}^{p}\bb{\tau}\right]
\label{jaumannzdp2}
\end{equation}

A particular stress-rate important for the next arguments is the Truesdell rate $\bb{\dot{\tau}}^L:=\bb{\dot{\tau}}-\bb{l}\bb{\tau}-\bb{\tau}\bb{l}^T$, which corresponds to $\bb{z}(\bb{\tau},\bb{d}) = \bb{\tau}\bb{d}+\bb{d}\bb{\tau}$. In this case, $\bb{K}^L:\bb{d} = \bb{K}^J:\bb{d}-\left(\bb{\tau}\bb{d}+\bb{d}\bb{\tau}\right)$,
for any symmetric second order tensor $\bb{d}$ and
\begin{equation}
\bb{d}^{pL} = \left(\bb{K}^L\right)^{-1}\bb{K}^J:\bb{d}^p
= \widehat{\bb{d}}^p + \left(\bb{K}^L\right)^{-1}\left(\bb{\tau}\bb{l}^{pT}+\bb{l}^p\bb{\tau}\right)
\label{Truesdelldp}
\end{equation}

\section{Schmid law and Kirchhoff stress}

\noindent
Schmid\cite{Schmid1924} law states that a slip system becomes
active, or is yielding, if its resolved shear stress reaches a
critical value. Traditionally, this is formulated in terms of the
Cauchy stress as
\begin{equation}
\tau^{\alpha}:=\bb{\sigma}\cdot\bb{g}^{\alpha}  = \bb{\sigma}\cdot\bb{a}^{\alpha} \leq \tau^{\alpha}_{cr}
\label{schmida}
\end{equation}
with $\tau^{\alpha}_{cr}=\tau^{\alpha}_{cr}(x,t)$ representing the
critical resolved shear stress of slip system $\bb{g}^{\alpha}$. A
slight generalization of the above activation criterion will simplify
significantly the formalism\,; Schmid law will be restated in the
form
\begin{equation}
\tau^{\alpha}:=\bb{\tau}\cdot\bb{g}^{\alpha} = \bb{\tau}\cdot\bb{a}^{\alpha} \leq \tau^{\alpha}_{cr}
\label{schmidb}
\end{equation}
While formally similar, the two activation/yielding criteria
differ, the second being sensitive to the hydrostatic component of
the stress state. In support of the above generalization let us note
that:

\noindent \bb{i}) For isochoric motions, $J=constant$; then for
$J=1$ criterions (\ref{schmida}) and (\ref{schmidb}) are
equivalent.

\noindent \bb{ii}) Eq.(\ref{schmidb}) is equivalent to
$\bb{\sigma}\cdot\bb{g}^{\alpha}\leq\tau^{\alpha}_{cr}/J$.
Assuming there is a one-to-one relationship $J=J(p)$, where
$p:=\tr(\bb{\sigma})/3$, this is necessarily a strictly increasing
function. Then adding to an existing yielding stress state a
dilatational pressure component will cause further yielding while
a compressive pressure component will cause elastic unloading.
This is in qualitative agreement with the experiments of
Spitzig and Richmond\cite{SR84}. However, it should be noted that this argument
applies rigorously only to crystals featuring cubic or a
particular form of hexagonal symmetry, in which case the elastic
properties of the crystal admit a volumetric-deviatoric
decomposition with a one-to-one relationship $J=J(p)$. 

\noindent \bb{iii}) Employing a Taylor series expansion about
$p=0$ and assuming that $J(0) = 1$, eq.(\ref{schmidb}) can be
rewritten as
$\bb{\sigma}\cdot\bb{g}^{\alpha}\leq\tau^{\alpha}_{cr}\left[1-qp+\mathscr{O}(p^2)\right]$,
where $q:=J^\prime(0)$. For small displacements from the reference
configuration, and thus in a small neighborhood of the zero stress
state, Hooke's law for cubic crystals can be expressed as
$\bb{\sigma} =
(1/3)(c_{11}+2c_{12})\tr(\bb{\epsilon})\bb{I}+$deviatoric terms,
\cite{MCowin}, where $c_{ij}$ are the components of the tensor of
elasticity in Voigt notation, $\bb{\epsilon}$ is the strain tensor
associated with small displacements, and $\bb{I}$ is the second
order identity tensor. In this case $p = \tr(\bb{\sigma})/3 =
(c_{11}+2c_{12})\tr(\bb{\epsilon})/3$ and hence $J =
1+\tr(\bb{\epsilon})+\mathscr{O}(\bb{\epsilon}^2) =
1+3p/(c_{11}+2c_{12})+\mathscr{O}(p^2)$, allowing us to deduce
that $q = J^\prime(0) = 3/(c_{11}+2c_{12})$. On the other hand, it
will be shown elsewhere that a realistic extension of Schmid law,
incorporating pressure effects in accordance with the experiments
of \cite{SR84}, is of the form
$\bb{\sigma}\cdot\bb{g}^{\alpha}\leq
\tau^{\alpha}_{cr}\left(1-3\alpha p\right)$, where $\alpha\approx
56/$TPa for aluminum and in the range $13-23/$TPa for iron and
steels. For aluminum, $c_{11}\approx 107$ GPa and $c_{12}\approx
61$ GPa, leading to $q/3\approx 4/$TPa, which is about sixteen
times less than $\alpha$\,; for iron, $c_{11}\approx 237$ GPa and
$c_{12}\approx 141$ GPa, leading to $q/3\approx 2/$TPa, which is
about ten times less than observed in experiments.

\noindent
\bb{iv})
Finally, consider a purely hydrostatic stress state, $\bb{\sigma} = p\bb{I}$.
Then $\tau^{\alpha} = Jp\bb{I}\cdot\bb{g}^{\alpha} = 0$, since $\bb{m}^{\alpha}\cdot\bb{n}^{\alpha} = 0$,
and hence the strict inequality $0=\tau^{\alpha}<\tau^{\alpha}_{cr}$ is satisfied. Thus a purely
hydrostatic stress state will never cause the yielding of the crystal.

All of the above arguments indicate that eq.(\ref{schmidb}) is a
valid extension of the classical Schmid law, the pressure
component of the stress state having only a weak influence upon
slip system activation, yet being in qualitative agreement with a
more realistic extension of Schmid law to pressure effects.

\section{Elastic directions and normality}

\noindent When strict inequality holds in eq.(\ref{schmidb}) for
all slip systems, the stress state is elastic. The set of all
elastic stress states defines the (current) elastic domain at
particle $x$ and its boundary is called the activation surface.
The relationship between the current rate of plastic deformation
and the yield surface is crucial for the structure of the overall
response of an aggregate of single crystals. Characterizing this
relationship is equivalent to characterizing the directions of
elastic unloading or, in the terminology of Hill and Rice\cite{HillRice72},
equivalent to specifying the unloading criterion. This in turn is
equivalent to specifying how the slip systems evolve during
deformation.

There are many plausible definitions for the evolution of the
"slip" vectors $\bb{m}^{\alpha}$ and $\bb{n}^{\alpha}$,
\cite{HillRice72}, \cite{AsaroRice}. They are, in fact,
phenomenological entities (internal variables) intended to
identify, at crystal level, the basic modes of plastic deformation
resulting from the motion, at atomic level, of a large number of
dislocations through the crystal lattice. As such, there is a
certain freedom in stating the evolution laws for the slip
vectors, although, in principle, $\bb{m}^{\alpha}$ should be
pointing in a direction of slip and $\bb{n}^{\alpha}$ in the
direction normal to the slip plane. A general representation of
this evolution is
\begin{equation}
\bb{\dot{m}}^{\alpha} =
\bb{l}_A(\bb{l}^e,\bb{m}^{\alpha}):\bb{m}^{\alpha},\,\,\,\,\,\,\,
\bb{\dot{n}}^{\alpha} =
\bb{l}_B(\bb{l}^e,\bb{n}^{\alpha}):\bb{n}^{\alpha}
\label{slipevol}
\end{equation}
with $\bb{l}_A$ and $\bb{l}_B$ second order tensor functions. As
illustrations, let us mention: the case where slip systems rotate
rigidly at the spin of the lattice, where $\bb{l}_A = \bb{l}_B =
\bb{w}^e$; the case where the slip direction convects with the
motion of the lattice while the normal to the slip plane convects
with the motion of the reciprocal frame, or, for short, convecting
slip systems, where $\bb{l}_A = \bb{l}^e$ and $\bb{l}_B =
-\bb{l}^{eT}$; the case of unit convecting slip systems, where, by
contrast with the previous case, the slip and normal vectors
retain their initial lengths, thus $\bb{l}_A
=\bb{l}^e-\left[\left(\bb{d}^e:\bb{m}^{\alpha}\right)\cdot\bb{m}^{\alpha}\right]\bb{I}$
and $\bb{l}_B
=-\bb{l}^{eT}+\left[\left(\bb{d}^e:\bb{n}^{\alpha}\right)\cdot\bb{n}^{\alpha}\right]\bb{I}$.
These examples suggest the following representation for the tensors $\bb{l}_A$ and $\bb{l}_B$
characterizing slip system evolution:
\begin{equation}
\bb{l}_A = \bb{l}^e+\bb{G}_A:\bb{d}^e,\,\,\,\,\bb{l}_B = -\bb{l}^{eT}+\bb{G}_B:\bb{d}^e
\label{slipevol2}
\end{equation}
with $\bb{G}_A$ and $\bb{G}_B$ fourth order tensors which may depend on $\bb{m}^{\alpha}$ and $\bb{n}^{\alpha}$:
$\bb{G}_B = -\bb{G}_A = $ the fourth order identity tensor, for rigidly rotating slip systems;
$\bb{G}_A = \bb{G}_B = \bb{0}$, the fourth order null tensor, for convecting slip systems;
$\bb{G}_A = -\bb{I}\otimes\bb{m}^{\alpha}\otimes\bb{m}^{\alpha}$ and
$\bb{G}_B = \bb{I}\otimes\bb{n}^{\alpha}\otimes\bb{n}^{\alpha}$, for unit convecting slip systems.
Let us remark that, in order for the slip vectors to remain orthogonal during any motion,
the tensors $\bb{G}_A$ and $\bb{G}_B$ must satisfy:
\begin{equation}
\frac{d}{dt}\left(\bb{m}^{\alpha}\cdot\bb{n}^{\alpha}\right) \equiv 0\,\, \iff\,\,
\left(\bb{G}_A^T:\bb{m}^{\alpha}\otimes\bb{n}^{\alpha}+\bb{G}_B^T:\bb{n}^{\alpha}\otimes\bb{m}^{\alpha}\right)
\cdot\bb{d}^e\equiv 0
\label{slipevol2b}
\end{equation}
It is assumed in what follows that the functions $\bb{l}_A$ and
$\bb{l}_B$ comply with eqs.(\ref{slipevol2}) and
(\ref{slipevol2b}).

The elastic directions, in stress space, can now be characterized
by considering elastic continuations of the motion from the
current state. Since unloading directions are of primary interest,
let the stress state $\bb{\tau}$ at the current moment $t$ be a
yielding state; thus at least one slip system is active,
eq.(\ref{schmidb}) being satisfied with the equality sign. Let
next $\bb{\tau}(t+\theta)$, $\theta\in(0,\theta_M]$, $\theta_M>0$, denote a curve
in stress space originating at $\bb{\tau}(t)$ and subject to the
requirement that all its points are elastic states (for
$\theta>0$), the curve being otherwise arbitrary. Following a
notation in Hill and Rice\cite{HillRice73}, the tangent of an unloading curve
at its origin will be denoted by
\begin{equation}
\delta\bb{\tau}:= \left.\frac{d}{d\theta}\bb{\tau}(t+\theta)\right|_{\theta=0} = \lim_{\theta\rightarrow 0} \frac{1}{\theta}\left[\bb{\tau}(t+\theta)-\bb{\tau}(t)\right]
\label{deltatau}
\end{equation}
and will be referred to as an unloading direction;
the elastic directions (rates) of other entities will also be specified by the $\delta$ symbol.
More generally, the notation will be employed to denote arbitrary variations of an object with respect to the current configuration, that is,
the operation of derivation defined in eq.(\ref{deltatau}).

We show next a fundamental invariance property relating elastic
directions and the direction of plastic flow. But first, some remarks
regarding the continuity of the passage from plastic to elastic
regime are in order. At each inner point of the crystal the stress
is time-continuous so that $\bb{\tau}(t) =
\bb{\tau}(t+):=\lim_{\theta\rightarrow 0}\bb{\tau}(t+\theta)$. On
the other hand, $\bb{l}^p$ will in general experience a
discontinuity (in time), $\bb{l}^p(t)\neq\bb{0}=\bb{l}^p(t+)$,
since for elastic unloading $\bb{l}^p(t+\theta) = \bb{0}$, for any
$\theta>0$. Then $\bb{l}^e(t+)=\bb{l}(t+)$ and further,
$\bb{w}^e(t+)=\bb{w}(t+)$. The rate at which slip systems evolve
will also experience a discontinuity at the elastic-plastic
transition. However, given eq.(\ref{slipevol}), their motion is
continuous, so that $\bb{g}^{\alpha}(t+)=\bb{g}^{\alpha}(t)$. To
simplify notation, we shall continue to write, for example,
$\bb{l}(t)$, or simply $\bb{l}$, instead of $\bb{l}(t+)$, the
appropriate value being clear from the very specific context of
elastic unloading. With these in mind, we extend the
$\delta$-notation and define
\begin{equation}
\delta^J\bb{\tau}:=\delta\bb{\tau}+\bb{\tau}\bb{w}-\bb{w}\bb{\tau},\,\,\,\,
\delta^L\bb{\tau}:=\delta\bb{\tau}-\bb{\tau}\bb{l}^T-\bb{l}\bb{\tau},\,\,\,\,
\delta^Z\bb{\tau}:=\delta^J\bb{\tau}-\bb{z}(\bb{\tau},\bb{d})
\label{deltarates}
\end{equation}
For elastic unloading, from eqs.(\ref{jaumann}) and (\ref{jaumannz}) there holds
$\delta^J\bb{\tau} = \bb{K}^J:\bb{d}$ and  $\delta^Z\bb{\tau} = \bb{K}^Z:\bb{d}$, allowing us to
unfold the following sequence of equalities:
\begin{equation}
\begin{array}{c}
\dd\delta^Z\bb{\tau}\cdot\bb{d}^{pZ} = \left(\bb{K}^Z:\bb{d}\right)\cdot\bb{d}^{pZ} =
\left[\bb{K}^Z\left(\bb{K}^J\right)^{-1}:\delta^J\bb{\tau}\right]\cdot\bb{d}^{pZ} =\vp\\
\dd=\delta^J\bb{\tau}\cdot\left[\left(\bb{K}^J\right)^{-1}\bb{K}^Z:\bb{d}^{pZ}\right]
\end{array}
\end{equation}
where the symmetry of $\bb{K}^J$ and $\bb{K}^Z$ has been employed. With eq.(\ref{jaumannzdp})
we deduce
\begin{equation}
\delta^Z\bb{\tau}\cdot\bb{d}^{pZ} = \delta^J\bb{\tau}\cdot\bb{d}^p
\label{invariance}
\end{equation}
This is the announced invariance property. Hill\cite{Hill68b} was the
first to state it explicitly and to recognize its importance for
the foundations of the theory, \cite{HillRice72},
\cite{HillRice73}. Indeed, since the identity in
eq.(\ref{invariance}) holds for any elastic direction, it follows
that the relationship between the rate of plastic deformation and
the activation surface is frame-independent, and, moreover, in the
wider sense employed here for this concept, independent of
stress-strain measure or reference configuration.

At any moment along an elastic trajectory there holds
\begin{equation}
\bb{\tau}(t+\theta)\cdot\bb{a}^{\alpha}(t+\theta)<\tau^{\alpha}_{cr}(t) = \bb{\tau}(t)\cdot\bb{a}^{\alpha}(t), \,\,\,\,
\forall \alpha\in\mathcal{A}
\label{schmidelastic}
\end{equation}
Subtracting the right-hand member, dividing by $\theta$ and then taking the limit $\theta\rightarrow 0$ obtains
\begin{equation}
\delta\left(\bb{\tau}\cdot\bb{a}^{\alpha}\right)\leq 0 \,\,\iff\,\,
\delta\bb{\tau}\cdot\bb{a}^{\alpha}(t)+\bb{\tau}(t)\cdot\delta\bb{a}^{\alpha}\leq 0,\,\forall\alpha\in\mathcal{A}
\label{schmidelastic2}
\end{equation}
This inequality characterizes the elastic directions;
recalling the definition of the slip tensor $\bb{a}^{\alpha}$ in eq.(\ref{ratedef2}) and that of its evolution in eq.(\ref{slipevol}),
it becomes
\begin{equation}
\delta\bb{\tau}\cdot\bb{a}^{\alpha}+\frac{1}{2}\bb{\tau}\cdot
\left(\bb{l}_A\bb{g}^{\alpha}+\bb{g}^{\alpha T}\bb{l}_A^{T}+\bb{l}_B\bb{g}^{\alpha T}+\bb{g}^{\alpha}\bb{l}_B^T\right)\leq 0
\label{schmidelastic3}
\end{equation}
Let us denote for the moment
\begin{displaymath}
\bb{\mu}^{Z\alpha}:=\bb{a}^{\alpha}+\left(\bb{K}^Z\right)^{-1}:
\left[\bb{z}(\bb{\tau},\bb{a}^{\alpha})-\bb{\tau}(\bb{g}^{\alpha}-\bb{g}^{\alpha T})/2+(\bb{g}^{\alpha}-\bb{g}^{\alpha T})\bb{\tau}/2\right];
\end{displaymath}
by adding $0=\delta^Z\bb{\tau}\cdot\bb{\mu}^{Z\alpha}-\delta^Z\bb{\tau}\cdot\bb{\mu}^{Z\alpha}$,
inequality (\ref{schmidelastic3}) can be further reformulated as
\begin{equation}
\begin{array}{l}
\dd\delta^Z\bb{\tau}\cdot\bb{\mu}^{Z\alpha}+
\left(\bb{w}\bb{\tau}-\bb{\tau}\bb{w}+\bb{\tau}\bb{d}-\bb{d}\bb{\tau}+\bb{l}^T_A\bb{\tau}+\bb{\tau}\bb{l}_B\right)\cdot\bb{g}^{\alpha}+\vp\\
\dd+\bb{z}(\bb{\tau},\bb{d})\cdot\bb{a}^{\alpha}-\bb{z}(\bb{\tau},\bb{a}^{\alpha})\cdot\bb{d}\leq 0
\end{array}
\end{equation}
By the symmetry of $\bb{z}$ there holds $\bb{z}(\bb{\tau},\bb{d})\cdot\bb{a}^{\alpha}=\bb{z}(\bb{\tau},\bb{a}^{\alpha})\cdot\bb{d}$;
then, by multiplying the above inequality with $\dot{\gamma}^{\alpha}$, then summing for $\alpha\in\mathcal{A}$
and taking into account the linearity of $\bb{z}$ with respect to its second argument, one finally obtains the inequality
\begin{equation}
\delta^Z\bb{\tau}\cdot\bb{d}^{pZ}+\psi\leq 0,\,\,\,\,\makebox{with}\,\,\,\,\psi:=\sum_{\alpha\in\mathcal{A}}\dot{\gamma}^{\alpha}
\left[\bb{\tau}\left(\bb{l}_B+\bb{l}^T\right)+\left(\bb{l}_A^T-\bb{l}^T\right)\bb{\tau}\right]\cdot\bb{g}^{\alpha}
\label{flowrule}
\end{equation}
Since it holds for arbitrary elastic directions
$\delta^Z\bb{\tau}$, it defines the orientation of the current
rate of plastic deformation $\bb{d}^{pZ}$ with respect to the
tangent cone at the activation surface at the current stress
state. One may also note that $\psi$ is independent of the stress-rate employed.

When $\psi\equiv0$, the above inequality reduces to
\begin{equation}
\delta^Z\bb{\tau}\cdot\bb{d}^{pZ}\leq 0
\label{flowrule2}
\end{equation}
which is the classical case of "normality". Obviously, this happens
for arbitrary deformations if and only if $\bb{l}_A = \bb{l}$ and $\bb{l}_B=-\bb{l}^T$, that is, if it is assumed
that the \emph{slip systems convect with the lattice}.

Any other type of slip system evolution induces deviations from
the normality structure. As illustration, consider the case where
slip systems are assumed to rotate rigidly at the lattice spin;
substituting $\bb{l}_A=\bb{l}_B=\bb{w}$ in the definition of
$\psi$ in eq.(\ref{flowrule}) and employing the definition of
$\bb{w}^p$ in eq.(\ref{ratedef2}), straightforward calculations
lead to
\begin{equation}
\psi = \left(\bb{\tau}\bb{w}^p-\bb{w}^p\bb{\tau}\right)\cdot\bb{d} = \left[\left(\bb{K}^Z\right)^{-1}:\left(\bb{\tau}\bb{w}^p-\bb{w}^p\bb{\tau}\right)\right]\cdot\delta^Z\bb{\tau}
\label{psirigid}
\end{equation}
Substituting this into eq.(\ref{flowrule}) and employing the definition of $\bb{d}^{pZ}$ in eq.(\ref{jaumannzdp2}) results in the inequality
\begin{equation}
\begin{array}{c}
\dd\delta^Z\bb{\tau}\cdot\left[\bb{d}^{pZ}+\left(\bb{K}^Z\right)^{-1}:\left(\bb{\tau}\bb{w}^p-\bb{w}^p\bb{\tau}\right)\right] =\vp\\
\dd = \delta^Z\bb{\tau}\cdot\left[\widehat{\bb{d}}^p+\left(\bb{K}^Z\right)^{-1}:\bb{z}(\bb{\tau},\widehat{\bb{d}}^p)\right]\leq 0
\end{array}
\label{flowrule3rigid}
\end{equation}
When $\bb{z}\equiv\bb{0}$, that is, when Jaumann stress-rate is employed, the second form of the above inequality
simplifies to
\begin{equation}
\delta^J\bb{\tau}\cdot\left[\bb{d}^{p}+\left(\bb{K}^J\right)^{-1}:\left(\bb{\tau}\bb{w}^p-\bb{w}^p\bb{\tau}\right)\right] =
\delta^J\bb{\tau}\cdot\widehat{\bb{d}}^p\leq 0
\label{flowrule3rigid2}
\end{equation}
Obviously, the second form of the inequality above could have been
obtained more simply by employing a direct argument.

Returning to the general case, where slip system evolution is characterized by eqs.(\ref{slipevol}) and (\ref{slipevol2}),
let us note here for later reference that, by following similar calculations as above, the deviation $\psi$ admits the representation
\begin{equation}
\begin{array}{c}
\dd\psi =\bb{d}\cdot\bb{\Psi} = \delta^Z\bb{\tau}\cdot\left[\left(\bb{K}^Z\right)^{-1}:\bb{\Psi}\right],\vp\\
\dd
\makebox{where}\,\,\,\,\,
\bb{\Psi}:=
\sum_{\alpha\in\mathcal{A}}\dot{\gamma}^{\alpha}
\left[\bb{G}_A^T:\left(\bb{\tau}\bb{g}^{\alpha T}\right)+\bb{G}_B^T:\left(\bb{\tau}\bb{g}^{\alpha}\right)\right]
\end{array}
\label{psi}
\end{equation}

\section{Overall characteristics: Yield surface and flow rule}

\noindent Let $\overline{X}$ denote a material particle of a
continuum body; at $\overline{X}$ the constitutive response is to
be defined as the average response of a representative volume
element (RVE), see \cite{Hill72} for a general context to the
averaging problem. The RVE is viewed as an aggregate of single
crystals, a polycrystal, each constituent crystal being defined by
its orientation and by its constitutive response as described in
the previous sections.
The motion of the body, as well as that of the aggregate
represented by $\overline{X}$, is referred to a global (fixed)
orthonormal frame; $\Omega_t$ will denote the domain occupied by
the polycrystal in its current (deformed) configuration and
$\Omega_0$ the domain occupied in some reference configuration. As
first homogenization principle, it is assumed that the boundary of
the aggregate is subject to the homogeneous displacement field
associated with the macroscopic deformation gradient at particle
$\overline{X}$. Thus, if $\overline{\bb{F}}$ designates the later
and $x=x(X,t)$ describes the motion of the aggregate, with $X$
labeling particles in $\Omega_0$ and $x$ specifying their position
at $t$, then $x(X,t)=\overline{\bb{F}}(t):X$, for all
$X\in\partial\Omega_0$, at any moment $t$. Differentiating with
respect to $t$ and defining the macroscopic velocity gradient by
$\bb{L}:=\dot{\overline{\bb{F}}}\left(\overline{\bb{F}}\right)^{-1}$,
the boundary condition acquires the more convenient form where the
velocity field is prescribed on the boundary by
$\bb{v}(x,t)=\bb{L}(t):x$, for all $x\in\partial\Omega_t$. $\bb{D}
= \left(\bb{L}+\bb{L}^T\right)/2$ and $\bb{W} =
\left(\bb{L}-\bb{L}^T\right)/2$ will denote the macroscopic rate
of deformation and spin, respectively.

With $|\Omega_0|$ and $|\Omega_t|$ denoting the volumes of the referential and current domains, let us define
\begin{equation}
\overline{J}:=\frac{|\Omega_t|}{|\Omega_0|} = \frac{1}{|\Omega_0|}\int_{\Omega_0}J(X,t)\,dX \,\,\Longrightarrow\,\,
\overline{J} = \det\left(\overline{\bb{F}}\right)
\end{equation}
The implication follows by showing first that
$(d/dt)\,{\overline{J}} = \overline{J}\,\tr(\bb{L})$, via the
boundary condition and the divergence theorem; since
$\det\left(\overline{\bb{F}}\right)$ satisfies the same ordinary
differential equation and
$\overline{J}(t=0)=1=\det\left[\overline{\bb{F}}(t=0)\right]$, the
two coincide.

Neglecting inertial terms, equilibrium of the aggregate is
equivalent with div$(\bb{\sigma})=0$, that is
$\partial\sigma_{ij}/\partial x_j \equiv 0$. As second
homogenization principle, it is assumed that the macroscopic
Cauchy stress $\bb{\Sigma}$ at particle $\overline{X}$ is the
direct average of the Cauchy stress field within the aggregate
\begin{equation}
\bb{\Sigma}=\frac{1}{|\Omega_t|}\int_{\Omega_t}\bb{\sigma}(x,t)\,dx\,\,\iff\,\,
\bb{T}:=\overline{J}\,\bb{\Sigma} = \frac{1}{|\Omega_0|}\int_{\Omega_0}\bb{\tau}\,dX
\end{equation}
the second equivalent form of this axiom defining the macroscopic
Kirchhoff stress $\bb{T}$. Then, by employing the equilibrium
equation and the boundary condition (in rate form), a textbook
calculation proves the following averaging formula
\begin{equation}
\bb{\dot{T}}^L:=\bb{\dot{T}}-\bb{L}\bb{T}-\bb{T}\bb{L}^T = \frac{\overline{J}}{|\Omega_t|}\int_{\Omega_t}\frac{1}{J}\,\bb{\dot{\tau}}^L\,dx
\label{rateaverage}
\end{equation}

Let us also recall Hill's Lemma: for an arbitrary symmetric second
order equilibrated tensor field $\bb{\sigma}^*$, thus satisfying
div$(\bb{\sigma}^*)=0$, and for an arbitrary second order
compatible field $\bb{l}^*$ homogeneous at the boundary, thus
deriving from a vector potential $\bb{v}^*$,
$\bb{l}^*\equiv\nabla\bb{v}^*$, with the field $\bb{v}^*$
satisfying $\bb{v}^*(x)=\bb{L}^*:x$, for $x\in\partial\Omega_t$,
there holds
\begin{equation}
\bb{\Sigma}^*\cdot \bb{L}^* = \frac{1}{|\Omega_t|}\int_{\Omega_t}\bb{\sigma}^*(x)\cdot\bb{l}^*(x)\,dx, \,\,\,\,
\makebox{where}\,\,\,\, \bb{\Sigma}^*:=\frac{1}{|\Omega_t|}\int_{\Omega_t}\bb{\sigma}^*dx
\label{HillLemma}
\end{equation}

Finally, regarding the range of validity of the above averaging
formulas, it is required that the velocity field be piecewise
$C^1$ (thus $\bb{v}$ be continuous and its spatial gradient
piecewise continuous) and that the stress field (and its rate) be
of class $C^1$ within each constituent. Stress and velocity
gradient discontinuities may be present along inner boundaries
between the constituent grains (provided equilibrium is enforced
at each boundary).

\subsection{Instantaneous macro-elastic response}
\label{SectionMacroElastic}

\noindent
The macroscopic elastic directions (in stress space) play at macroscopic level a similar role
with that played by the elastic directions at crystal (or local) level.
Therefore it is essential to have a precise characterization of the instantaneous
elastic response at a macroscopic particle $\overline{X}$ and of its relationship with the
corresponding elastic fields developed within the aggregate represented by $\overline{X}$.

Let $t$ be the current moment of the motion $x=x(X,t)$ and let $\bb{v}^{\delta}=\bb{v}^{\delta}(x)$, $x\in\Omega_t$,
denote an arbitrary variation of the motion about the current configuration:
\begin{equation}
x^{\theta}:=x(X,t+\theta):=x(X,t)+\theta\,\bb{v}^{\delta}(x)+\mathscr{O}(\theta^2)
\label{variationmotion}
\end{equation}
Obviously, $\bb{v}^{\delta}$ is the velocity of
the varied motion $x^{\delta}$ at the moment $t+0$, that is, $\delta x=\bb{v}^{\delta}$.
With respect to the reference configuration $\Omega_0$, the gradient of the varied motion is
$\bb{F}^{\theta} = \partial x^{\theta}/\partial X = \left(\bb{I}+\theta\,\nabla\bb{v}^{\delta}\right)\bb{F}+\mathscr{O}(\theta^2)$,
where $\bb{F} = (\partial x/\partial X)(X,t)$, and hence
\begin{equation}
\delta\bb{F} = \left.\frac{d}{d\theta}\bb{F}^{\theta}\right|_{\theta=0} = \bb{l}^{\delta}\bb{F}(X,t),
\,\,\,\,\makebox{with}\,\,\,\,\bb{l}^{\delta}:=\nabla\bb{v}^{\delta}=\frac{\partial\bb{v}^{\delta}}{\partial x}
\label{deltaF}
\end{equation}

Let $\bb{\sigma}^{\theta}:=\bb{\sigma}(x^{\theta},t+\theta)$ be the Cauchy stress associated with the motion $x^{\theta}$.
Neglecting inertial terms, an arbitrary part $P^{\theta} = x(P,t+\theta)$, $P\subseteq\Omega_0$, is in equilibrium if
at any $\theta\geq 0$ there holds
\begin{displaymath}
\int_{\partial P^{\theta}}\bb{\sigma}^{\theta}:\bb{n}^{\theta}d\Gamma^{\theta} = \bb{0}
\end{displaymath}
where $\bb{n}^{\theta}d\Gamma^{\theta}$ is the area element on the boundary $\partial P^{\theta}$.
This is related, by Nanson's formula, to the area element in the reference state by
$\bb{n}^{\theta}d\Gamma^{\theta} = J^{\theta}\left(\bb{F}^{\theta}\right)^{-T}:\bb{N}d\Gamma$;
then, by taking the $\theta$-rate at the current moment there follows:
\begin{displaymath}
\frac{d}{d\theta}\left(\int_{\partial P^{\theta}}\bb{\sigma}^{\theta}:\bb{n}^{\theta}d\Gamma^{\theta}\right)_{\theta=0} = \bb{0}\iff
\int_{\partial P} \delta\left(J\bb{\sigma}\bb{F}^{-T}\right):\bb{N}\,d\Gamma = \bb{0}
\end{displaymath}
With $J^{\theta}=\det(\bb{F}^{\theta})$, there follows $\delta J = J\tr(\bb{l}^{\delta})$;
from $\delta(\bb{F}\bb{F}^{-1}) = \bb{0}$ and eq.(\ref{deltaF}) there follows $\delta\bb{F}^{-T} = -\bb{l}^{\delta T}\bb{F}^{-T}$.
Employing these formulas and then once again Nanson's formula into the right-hand member of the above equivalence,
the stress variation induced by the variation of the motion (\ref{variationmotion}) satisfies the equilibrium equation:
\begin{displaymath}
\int_{\partial P_t}\frac{1}{J}\left(\delta^L\bb{\tau}+\bb{l}^{\delta}\bb{\tau}\right):\bb{n}\,d\Gamma_t = \bb{0},
\,\,\,\forall P_t\subseteq\Omega_t
\end{displaymath}
where, here, $\delta^L\bb{\tau} = \delta\bb{\tau}-\bb{l}^{\delta}\bb{\tau}-\bb{\tau}\bb{l}^{\delta T}$. 
By a classic argument, this is further equivalent to
\begin{displaymath}
\makebox{div}\left[\frac{1}{J}\left(\delta^L\bb{\tau}+\bb{l}^{\delta}\bb{\tau}\right)\right] = \bb{0},\,\,\,\forall x\in\Omega_t
\end{displaymath}

Let now $\overline{\bb{F}}^{\theta}:=\overline{\bb{F}}(t)+\theta\,\bb{L}^{\delta}+\mathscr{O}(\theta^2)$
represent a variation of the macroscopic deformation gradient in a direction $\bb{L}^{\delta}$. Via the boundary condition,
this variation will generate a variation of the motion as in eq.(\ref{variationmotion}).
If this motion induces an elastic deformation within the aggregate, for $\theta$ in some nonempty interval $(0,\theta_M]$,
then, considering all of the above, and taking the limit $\theta\rightarrow 0$,
the instantaneous response of the aggregate is characterized
by the following boundary value problem (BVP):
\begin{equation}
\left\{
\begin{array}{l}
\dd\makebox{div}\left[\frac{1}{J}\left(\delta^L\bb{\tau}+\bb{l}^{\delta}\bb{\tau}\right)\right]= \bb{0},\, x\in\Omega_t\vp\\
\dd\delta^L\bb{\tau} = \bb{K}^L:\bb{d}^{\delta},\,x\in\Omega_t,\,\,\,\makebox{where}\,\,\, \bb{d}^{\delta}:=(\bb{l}^{\delta}+\bb{l}^{\delta T})/2\vp\\
\dd\bb{v}^{\delta}(x) =
\bb{L}^{\delta}:x,\,\,\,x\in\partial\Omega_t
\end{array}\right.
\label{BVPelastic}
\end{equation}
and $\bb{L}^{\delta}$ will be called a macro-elastic direction of the motion.
Note that $J$, $\bb{\tau}$ and $\bb{K}^L$ are evaluated at the current moment $t$; then substituting
the instantaneous stress-strain response into the equilibrium equation, the following BVP
\begin{equation}
\left\{
\begin{array}{l}
\dd\makebox{div}\left[\frac{1}{J}\left(\bb{K}^L:\bb{d}^{\delta}+\bb{l}^{\delta}\bb{\tau}\right)\right]= \bb{0},\, x\in\Omega_t\vp\\
\dd\bb{v}^{\delta}(x) =
\bb{L}^{\delta}:x,\,\,\,x\in\partial\Omega_t
\end{array}\right.
\label{BVPelastic2}
\end{equation}
is \emph{linear} with respect to $\bb{L}^{\delta}$. Then a field
of linear operators $\bb{V}=\bb{V}(x)$ exists such that
$\bb{v}^{\delta}(x) = \bb{V}(x):\bb{L}^{\delta}$\, $\iff$\,
$v^{\delta}_i(x) = V_{iab}(x)L^{\delta}_{ab}$. In particular, if
$\bb{L}^{\delta}=\bb{W}^*$, with $\bb{W}^*$ an arbitrary
antisymmetric second order tensor, it can be verified that the
solution of BVP(\ref{BVPelastic2}) is the uniform velocity field
$\bb{v}^*(x) = \bb{W}^*:x$, $x\in\Omega$. As such, $0 = d^*_{ij} =
(1/2)\left(V_{iab,j}+V_{jab,i}\right)W^*_{ab}$ and $W^*_{ij} =
w^*_{ij} = (1/2)\left(V_{iab,j}-V_{jab,i}\right)W^*_{ab}$ (a comma
indicating partial differentiation with respect to the variable it
precedes); by adding these two identities there follows: $W^*_{ij}
= V_{iab,j}W^*_{ij}$. Then the fourth order operator
$\overline{A}$ defined by $\overline{A}_{ijab}:=V_{iab,j}$ relates
the local velocity gradient to the boundary condition through
$\bb{l}^{\delta}(x) = \overline{\bb{A}}(x):\bb{L}^{\delta}$, and
has the additional property that $\overline{\bb{A}}(x):\bb{W}^* =
\bb{W}^*$, $\forall\bb{W}^*$ antisymmetric. Thus, for an arbitrary
$\bb{L}^{\delta}$, with
$\bb{D}^{\delta}:=(\bb{L}^{\delta}+\bb{L}^{\delta T})/2$ and
$\bb{W}^{\delta}:=(\bb{L}^{\delta}-\bb{L}^{\delta T})/2$, the
solution of BVP(\ref{BVPelastic2}) can be described as follows:
\begin{equation}
\bb{d}^{\delta}(x) = \bb{A}^S(x):\bb{D}^{\delta},
\,\,\,x\in\Omega_t
\label{ASlocalization}
\end{equation}
where $A^S_{ijab}:=(\overline{A}_{ijab}+\overline{A}_{jiab})/2$, and
\begin{equation}
\bb{w}^{\delta}(x):= (\bb{l}^{\delta}-\bb{l}^{\delta T})/2 =
\bb{W}^{\delta}+\bb{A}^A(x):\bb{D}^{\delta}, \,\,\,x\in\Omega_t
\label{AAlocalization}
\end{equation}
where $A^A_{ijab}:=(\overline{A}_{ijab}-\overline{A}_{jiab})/2$.
$\bb{A}^S$ and $\bb{A}^A$ will be referred to as the localization
operators of the above BVP's. They are completely determined by
the current state of the aggregate (geometry, elastic properties
and stress state), via $\Omega_t$, $\bb{\tau}$ and $\bb{K}^L$.

The instantaneous macroscopic elastic response can now be deduced
by
\begin{displaymath}
\begin{array}{c}
\dd\frac{1}{J}\delta^L\bb{\tau}=\frac{1}{J}\bb{K}^L:\bb{d}^{\delta} = \frac{1}{J}\bb{K}^L\bb{A}^S:\bb{D}^{\delta}\Longrightarrow\vp\\
\dd\Longrightarrow\frac{\overline{J}}{|\Omega_t|}\int_{\Omega_t}\frac{1}{J}\delta^L\bb{\tau}\,dx =
\frac{\overline{J}}{|\Omega_t|}\int_{\Omega_t}\frac{1}{J}\bb{K}^L\bb{A}^S\,dx :\bb{D}^{\delta}
\end{array}
\end{displaymath}
allowing us to write, by recalling eq.(\ref{rateaverage}),
\begin{equation}
\begin{array}{c}
\dd\delta^L\bb{T} = \overline{\bb{K}}^L:\bb{D}^{\delta},\vp\\
\dd\makebox{with}\,\,\,\,
\delta^L\bb{T}:=\delta\bb{T}-\bb{L}^{\delta}\bb{T}-\bb{T}\bb{L}^{\delta T}\,\,\,\,\makebox{and}\,\,\,\,
\overline{\bb{K}}^L:=\frac{\overline{J}}{|\Omega_t|}\int_{\Omega_t}\frac{1}{J}\bb{K}^L\bb{A}^S\,dx
\end{array}
\label{KLMacro}
\end{equation}
$\overline{\bb{K}}^L$ is the macroscopic tensor of elasticity;
crucial for the next developments, $\overline{\bb{K}}^L$ is
symmetric and positive definite, once $\bb{K}^L$ is so. The proofs
are somewhat technical and therefore are detailed in the Appendix
to this work. In particular, $\overline{\bb{K}}^L$ is invertible
and then $\delta^{L}\bb{\tau} = \bb{K}^L\bb{A}^S:\bb{D}^{\delta} =
\bb{K}^L\bb{A}^S\left(\overline{\bb{K}}^L\right)^{-1}:\delta^L\bb{T}$,
this representing the relationship between the elastic variations
of the local and macroscopic stresses:
\begin{equation}
\delta^{L}\bb{\tau}(x) = \bb{B}^L(x):\delta^L\bb{T},\,\,\,\,\makebox{with}\,\,\,\,\bb{B}^L:=\bb{K}^L\bb{A}^S\left(\overline{\bb{K}}^L\right)^{-1}
\label{BTensor}
\end{equation}

\subsection{Macro-rate of plastic deformation}

\noindent
Consider next a general deformation process subjecting the aggregate to elastic-plastic deformation.
Obviously, the actual velocity field at the current moment $t$ is among the possible directions of variation in eq.(\ref{variationmotion});
hence the BVP governing the evolution of the aggregate in this case reads
\begin{equation}
\left\{
\begin{array}{l}
\dd\makebox{div}\left[\frac{1}{J}\left(\bb{\dot{\tau}}^L+\bb{l}\bb{\tau}\right)\right]= \bb{0},\, x\in\Omega_t\vp\\
\dd\bb{\dot{\tau}}^L = \bb{K}^L:\left(\bb{d}-\bb{d}^{pL}\right),\,x\in\Omega_t\vp\\
\dd\bb{v}(x,t) = \bb{L}(t):x,\,\,\,x\in\partial\Omega_t
\end{array}\right.
\label{BVPplastic}
\end{equation}
To characterize the overall (or macroscopic) stress-strain response associated with the above BVP, we shall employ a technique essentially
due to Hill\cite{Hill67} and Hill\cite{Hill72};
in \cite{Hill67} the micro-macro transition was achieved at the cost of neglecting structural changes (texture evolution) generated by plastic deformation, whereas in \cite{Hill72} it was shown, under general conditions, that if the normality structure is present at crystal level
then it propagates intact at macroscopic level. An alternative construction of the theory
was provided by Hill and Rice\cite{HillRice73} but the approach is essentially based on the elastic potential and its parametrization.
Here, the theory developed in \cite{Hill67} is extended to a general Eulerian framework, without specification
of the constitutive origins of the instantaneous elastic response.
Nevertheless, as in the cited works,
the key to the micro-macro transition will be the differential invariant in eq.(\ref{invariance}).

Given the current state of the aggregate, represented by its domain $\Omega_t$ and the fields $\bb{\tau}$ and $\bb{K}^L$,
let $\bb{L}^{\delta}$ denote an arbitrary macro-elastic direction of motion, $\delta^L\bb{T} = \overline{\bb{K}}^L:\bb{D}^{\delta}$
its corresponding macro-elastic stress direction, and $\bb{l}^{\delta}$ and $\delta^L\bb{\tau}$ the corresponding
local fields associated with $\bb{L}^{\delta}$ via BVP(\ref{BVPelastic}). Then the following sequence of equalities unfolds:
\begin{equation}
\delta^L\bb{\tau}\cdot\bb{d}^{pL} = \delta^L\bb{\tau}\cdot\left(\bb{d}-\bb{d}^e\right) =
\delta^L\bb{\tau}\cdot\bb{d} - \left(\bb{K}^L:\bb{d}^{\delta}\right)\cdot\bb{d}^e =
\delta^L\bb{\tau}\cdot\bb{d} - \bb{\dot{\tau}}^L\cdot\bb{d}^{\delta}
\end{equation}
Observing now that $\left(\bb{l}^{\delta}\bb{\tau}\right)\cdot\bb{l} = \left(\bb{l}\bb{\tau}\right)\cdot\bb{l}^{\delta}$,
the equality between the first and the last members of the above sequence can be equivalently restated as
\begin{equation}
\frac{1}{J}\,\delta^L\bb{\tau}\cdot\bb{d}^{pL} =
\frac{1}{J}\left[\left(\delta^L\bb{\tau}+\bb{l}^{\delta}\bb{\tau}\right)\cdot\bb{l}-
\left(\bb{\dot{\tau}}^L+\bb{l}\bb{\tau}\right)\cdot\bb{l}^{\delta}\right]
\end{equation}
As solutions of BVP(\ref{BVPelastic}) and BVP(\ref{BVPplastic}), the fields $\delta^L\bb{\tau}+\bb{l}^{\delta}\bb{\tau}$ and  $\bb{\dot{\tau}}^L+\bb{l}\bb{\tau}$ are equilibrated, while $\bb{l}$ and $\bb{l}^{\delta}$ are compatible;
then taking the average of both members of the above equality and employing
Hill's Lemma (\ref{HillLemma}) and formula (\ref{rateaverage}), obtains
\begin{equation}
\frac{\overline{J}}{|\Omega_t|}\int_{\Omega_t}\frac{1}{J}\,\delta^L\bb{\tau}\cdot\bb{d}^{pL}dx =
\left(\delta^L\bb{T}+\bb{L}^{\delta}\bb{T}\right)\cdot\bb{L} - \left(\bb{\dot{T}}^L+\bb{L}\bb{T}\right)\cdot\bb{L}^{\delta}
\label{inva}
\end{equation}
Again, due to the symmetry of $\bb{T}$, there holds
$\left(\bb{L}^{\delta}\bb{T}\right)\cdot\bb{L} = \left(\bb{L}\bb{T}\right)\cdot\bb{L}^{\delta}$,
and then, by eq.(\ref{KLMacro}), the right-hand member of the above equality transforms into
\begin{equation}
\delta^L\bb{T}\cdot\bb{D} - \bb{\dot{T}}^L\cdot\bb{D}^{\delta}  = \left[\bb{D}-\left(\overline{\bb{K}}^L\right)^{-1}:\bb{\dot{T}}^L\right]\cdot\delta^L\bb{T}
\label{invb}
\end{equation}
due to the symmetry of $\delta^L\bb{T}$, $\bb{\dot{T}}^L$ and $\overline{\bb{K}}^L$.

By \emph{definition}, the bracketed term in eq.(\ref{invb}) is the
macroscopic rate of plastic deformation associated with the rate
$\bb{\dot{T}}^L$; it will be denoted by $\bb{D}^{pL}$. This is
equivalent with assuming that the overall stress-strain response
must be, in rate form:
\begin{equation}
\bb{\dot{T}}^L = \overline{\bb{K}}^{L}:\left(\bb{D}-\bb{D}^{pL}\right)
\label{macrorateT}
\end{equation}
Consequently, by eqs.(\ref{inva}) and (\ref{invb}),
for any macro-elastic direction $\delta^L\bb{T}$, the macro-rate of plastic deformation satisfies
\begin{equation}
\bb{D}^{pL}\cdot\delta^L\bb{T} = \frac{\overline{J}}{|\Omega_t|}\int_{\Omega_t}\frac{1}{J}\,\delta^L\bb{\tau}\cdot\bb{d}^{pL}dx \iff
\bb{D}^{pL} = \frac{\overline{J}}{|\Omega_t|}\int_{\Omega_t}\frac{1}{J}\,\left(\bb{B}^L\right)^T:\bb{d}^{pL}dx
\label{macrorateDP}
\end{equation}
Excluding pathological cases
, the set of macro-elastic directions has a non-empty interior and hence it contains six (the dimension of the space of symmetric second order tensors) linearly independent elements. Then the second equality in eq.(\ref{macrorateDP}) follows by employing eq.(\ref{BTensor}); it demonstrates that the macro-rate of plastic deformation enjoys an unequivocal representation in terms of
the plastic deformation taking place within the aggregate represented by the macroscopic particle $\overline{X}$.

With arguments similar to those in sections 2 and 4, once equations (\ref{macrorateT}) and (\ref{macrorateDP}) have been
formulated by one observer, they can be reformulated in terms of any reference frame.
For later reference we state here the main formulas. For the Jaumann rate $\bb{\dot{T}}^J:=\bb{\dot{T}}+\bb{T}\bb{W}-\bb{W}\bb{T}$,
the corresponding tensor of instantaneous moduli is defined by $\overline{\bb{K}}^J:\bb{A} = \overline{\bb{K}}^L:\bb{A}+\bb{T}\bb{A}+\bb{A}\bb{T}$
for any symmetric second order tensor $\bb{A}$, and is symmetric and positive definite; the corresponding stress-strain relationship
reads:
\begin{equation}
\bb{\dot{T}}^J = \overline{\bb{K}}^J:\left(\bb{D}-\bb{D}^p\right),\,\,\,\,
\makebox{with}\,\,\,\, \bb{D}^p:=\left(\overline{\bb{K}}^J\right)^{-1}\overline{\bb{K}}^L:\bb{D}^{pL}
\label{jaumannMacro}
\end{equation}
More generally, for an arbitrary stress-rate $\bb{\dot{T}}^Z:=\bb{\dot{T}}^J-\overline{\bb{Z}}(\bb{T},\bb{D})$
with $\overline{\bb{Z}}$ linear with respect to $\bb{D}$, the tensor $\overline{\bb{K}}^Z$ defined by $\overline{\bb{K}}^Z:\bb{A} = \overline{\bb{K}}^J:\bb{A}-\overline{\bb{Z}}(\bb{T},\bb{A})$,
for any symmetric second order tensor $\bb{A}$, is symmetric and positive definite and:
\begin{equation}
\bb{\dot{T}}^Z = \overline{\bb{K}}^Z:\left(\bb{D}-\bb{D}^{pZ}\right),\,\,\,\,
\makebox{with}\,\,\,\, \bb{D}^{pZ}:=\left(\overline{\bb{K}}^Z\right)^{-1}\overline{\bb{K}}^J:\bb{D}^{p}
\end{equation}
Finally, the following sequence of equalities holds, by virtue of the invariance property in eq.(\ref{invariance}), which is equally valid at macroscopic level, and by eq.(\ref{macrorateDP}):
\begin{equation}
\bb{D}^{pZ}\cdot\delta^Z\bb{T} = \bb{D}^{p}\cdot\delta^J\bb{T} = \bb{D}^{pL}\cdot\delta^L\bb{T} = \frac{\overline{J}}{|\Omega_t|}\int_{\Omega_t}\frac{1}{J}\,\delta^L\bb{\tau}\cdot\bb{d}^{pL}dx
\label{invarianceMacro}
\end{equation}

\subsection{Macroscopic yield surface}

\noindent
The macroscopic elastic domain is, by definition, the set of all stress states that can be attained
from the current stress by purely elastic deformation, \cite{Hill67};
the macroscopic yield surface is the boundary of the elastic domain.
Eq.(\ref{invarianceMacro}) shows the invariance of the relationship between the macroscopic rate of deformation
and the macroscopic yield surface. It is therefore sufficient to discuss the nature of this relationship
for one description and the conclusions will transfer unchanged to any other description.
In the present Eulerian framework, the most convenient description is that associated with the Jaumann rate.

Let $\bb{T}(t)$ denote the current macro-stress, and $\bb{\tau}(x,t)$ its corresponding stress field within the aggregate.
During any elastic deformation originating at $\bb{T}$ and driven by an evolution $\bb{L}=\bb{L}(t+\theta)$ at the boundary,
with $\bb{\dot{T}}^J(t+\theta) = \overline{\bb{K}}^J(t+\theta):\bb{D}(t+\theta)$,
for $\theta$ in some interval $\in(0,\theta^*]$,
the local stress field $\bb{\tau}(x,t+\theta)$ satisfies at any $t+\theta$
\begin{equation}
\bb{\dot{\tau}}^J = \bb{K}^J:\bb{d} = \bb{K}^J\bb{A}^S:\bb{D} = \bb{K}^J\bb{A}^S\left(\overline{\bb{K}}^J\right)^{-1}:\bb{\dot{T}}^J
=\bb{B}^J:\bb{\dot{T}}^J
\label{stress1}
\end{equation}
where derivatives are taken with respect to $\theta$ and the last
equality from the above sequence serves as definition of the
operator $\bb{B}^J$. In order to obtain a representation of the
local field $\bb{\tau}(t+\theta^*)$ in terms of
$\bb{T}^*:=\bb{T}(t+\theta^*)$, the localization operators
$\bb{A}^S$ and $\bb{A}^A$ in eqs.(\ref{ASlocalization}) and
(\ref{AAlocalization}) may be considered as known; the operator
$\bb{B}^J$ is then known, as well as the field $\bb{w} =
\bb{w}^i+\bb{W}$, where $\bb{w}^i:=\bb{A}^A:\bb{D}$,
eq.(\ref{AAlocalization}). With $\bb{W}$ we associate the
macro-rotation $\bb{R}$ satisfying $\bb{\dot{R}} = \bb{W}\bb{R}$;
with $\bb{w}^{iR}:=\bb{R}^T\bb{w}^i\bb{R}$ we associate the
rotation field $\bb{Q}(x,t+\theta)$ defined by $\bb{\dot{Q}} =
\bb{w}^{iR}\bb{Q}$ and $\bb{Q}(t) = \bb{I}$ (and note that
$\bb{Q}$ is completely determined by the initial condition and the
evolution of the boundary condition); we also define the so called rotated
stresses $\bb{T}^{R}:=\bb{R}^T\bb{T}\bb{R}$,
$\bb{T}^{RQ}:=\bb{Q}^T\bb{T}^R\bb{Q}$,
$\bb{\tau}^R:=\bb{R}^T\bb{\tau}\bb{R}$, and
$\bb{\tau}^{RQ}:=\bb{Q}^T\bb{\tau}^R\bb{Q}$, and note that, for
example, $\bb{\dot{T}}^J =
\bb{R}\left(d\bb{T}^R/d\theta\right)\bb{R}^T$. Under these
assumptions, eq.(\ref{stress1}) becomes an ordinary differential
equation, which, after employing the just defined rotated objects,
acquires the equivalent form
\begin{displaymath}
\frac{d}{d\theta}\bb{\tau}^{RQ} = \bb{Q}^T\bb{R}^T\left\{\bb{B}^J:\left[\bb{R}\left(\frac{d}{d\theta}\bb{T}^R\right)\bb{R}^T\right]\right\}\bb{R}\bb{Q}=
\bb{\Gamma}:\frac{d}{d\theta}\bb{T}^R
\end{displaymath}
with the second equality serving as definition for the fourth order operator $\bb{\Gamma}$.
Integrating between $\theta=0$ and $\theta=\theta^*$ there follows
\begin{displaymath}
\bb{\tau}^{RQ}(t+\theta^*)-\bb{\tau}^R(t) = \int_{0}^{\theta^*}\bb{\Gamma}:\,d\bb{T}^{R}
\end{displaymath}
In particular, the above integral vanishes for closed cycles (by
the assumed elastic behavior), and hence there exists a tensor
function $\bb{\Phi}=\bb{\Phi(\bb{T}^{R})}$ potential for its
integrand; then the stress field within the aggregate admits the
representation
\begin{equation}
\bb{\tau}^{RQ}(t+\theta^*)  =\bb{\tau}^{R}(t)+\bb{\Phi}(\bb{T}^R(t+\theta^*))-\bb{\Phi}(\bb{T}^R(t))
\label{stress2}
\end{equation}

Consider next the evolution of the slip systems of a constituent
during elastic unloading. In the context of Section 4, we shall
detail, as illustrations of a general argument, the cases of
rigidly rotating slip systems and of convecting slip systems. For
the first, the evolution of the slip direction $\bb{m}^{\alpha}$
and of the slip plane normal $\bb{n}^{\alpha}$ is of the type:
\begin{displaymath}
\frac{d}{d\theta}\bb{m}(t+\theta) = (\bb{w}^i+\bb{W})(t+\theta):\bb{m}(t+\theta)
\end{displaymath}
Recalling the definitions of the rotations $\bb{R}$ and $\bb{Q}$, these may be recognized as fundamental matrices of equations of the
above type. Then straightforward arguments lead to the representations
\begin{displaymath}
\begin{array}{c}
\dd\bb{m}^{\alpha}(t+\theta) = \bb{R}(t+\theta)\bb{Q}(t+\theta)\bb{R}^T(t):\bb{m}^{\alpha}(t)\vp\\
\dd\bb{n}^{\alpha}(t+\theta) = \bb{R}(t+\theta)\bb{Q}(t+\theta)\bb{R}^T(t):\bb{n}^{\alpha}(t)
\end{array}
\end{displaymath}
from where it follows that
\begin{equation}
\bb{a}^{\alpha}(t+\theta) = \left(\bb{RQ}\right)(t+\theta)\left[\bb{R}^T(t)\bb{a}^{\alpha}(t)\bb{R}(t)\right]\left(\bb{RQ}\right)^T(t+\theta)
\label{Arigid}
\end{equation}
Substituting the representations in eqs.(\ref{stress2}) and (\ref{Arigid}) into eq.(\ref{schmidelastic}),
one obtains the characterization of the current macroscopic elastic domain
as the set of all stress states $\bb{T}^*$ satisfying the set of inequalities
\begin{equation}
\bb{\tau}^{RQ}(t+\theta^*)\cdot\left(\bb{a}^{\alpha}\right)^R(t) =
\left[\bb{\tau}^{R}(t)-\bb{\Phi}(\bb{T}^R(t))+\bb{\Phi}(\bb{T}^{*R})\right]\cdot\left(\bb{g}^{\alpha}\right)^R(t)<\tau^{\alpha}_{cr}(t),
\label{elastic1}
\end{equation}
$\forall x\in\Omega_t$ and $\forall \alpha\in\mathcal{S}$.

In the case of convecting slip systems, the slip directions evolve during elastic unloading according to
\begin{equation}
\frac{d}{d\theta}\bb{m}(t+\theta) = \bb{l}(t+\theta):\bb{m}(t+\theta) =
\left(\bb{w}^i+\bb{W}+\bb{d}\right)(t+\theta):\bb{m}(t+\theta)
\label{slipconvect1}
\end{equation}
With the substitution $\bb{m}(t+\theta) = \bb{R}(t+\theta)\bb{Q}(t+\theta)\widetilde{\bb{m}}(t+\theta)$,
the rotational parts associated with the spins $\bb{w}^i$ and $\bb{W}$ are eliminated, $\widetilde{\bb{m}}$ satisfying
\begin{displaymath}
\frac{d}{d\theta}\widetilde{\bb{m}}(t+\theta) = \bb{d}^{RQ}(t+\theta):\widetilde{\bb{m}}(t+\theta)
\end{displaymath}
where
$\bb{d}^{RQ}:=\left(\bb{RQ}\right)^T\bb{d}\left(\bb{RQ}\right)$.
With $\bb{M}(t+\theta)$ denoting the fundamental matrix/operator
of the above equation, satisfying $\bb{M}(t)=\bb{I}$,
$\widetilde{\bb{m}}$ acquires the representation
$\widetilde{\bb{m}}(t+\theta) =
\bb{M}(t+\theta):\widetilde{\bb{m}}(t)$; the solution of
eq.(\ref{slipconvect1}) is then
\begin{equation}
\bb{m}(t+\theta) = \bb{R}(t+\theta)\bb{Q}(t+\theta)\bb{M}(t+\theta)\bb{R}^T(t):\bb{m}(t)
\label{slipconvect1b}
\end{equation}
On the other hand, normal directions evolve during elastic unloading according to
\begin{equation}
\frac{d}{d\theta}\bb{n}(t+\theta) =  -\bb{l}^T(t+\theta):\bb{n}(t+\theta) =
 \left(\bb{w}^i+\bb{W}-\bb{d}\right)(t+\theta):\bb{n}(t+\theta)
\label{slipconvect2}
\end{equation}
With $\bb{M}$ invertible, as fundamental operator, it can be verified that the solution of the above equation is
\begin{equation}
\bb{n}(t+\theta) = \bb{R}(t+\theta)\bb{Q}(t+\theta)\bb{M}^{-T}(t+\theta)\bb{R}^T(t):\bb{n}(t)
\label{slipconvect2b}
\end{equation}
Eqs. (\ref{slipconvect1b}) and (\ref{slipconvect2b}) then lead to
\begin{equation}
\bb{a}^{\alpha}(t+\theta) = \left(\bb{RQ}\right)(t+\theta)
\left[\widetilde{\bb{M}}(t+\theta):\left(\bb{g}^{\alpha}\right)^R(t)\right]\left(\bb{RQ}\right)^T(t+\theta)
\end{equation}
with the definition $\widetilde{\bb{M}}(t+\theta):\left(\bb{g}^{\alpha}\right)^R(t) := \bb{M}(t+\theta)\left(\bb{g}^{\alpha}\right)^R(t)\bb{M}^{-1}(t+\theta)
+\bb{M}^{-T}(t+\theta)\left[\left(\bb{g}^{\alpha}\right)^R(t)\right]^T\bb{M}^{T}(t+\theta)$.

Thus, in the case of convecting slip systems, the current macroscopic elastic domain
is the set of all stress states $\bb{T}^*$ satisfying the set of inequalities
\begin{equation}
\left\{\widetilde{\bb{M}}^T(t+\theta^*):
\left[\bb{\tau}^{R}(t)-\bb{\Phi}(\bb{T}^R(t))+\bb{\Phi}(\bb{T}^{*R})\right]\right\}\cdot\left(\bb{g}^{\alpha}\right)^R(t)<\tau^{\alpha}_{cr}(t),
\label{elastic2}
\end{equation}
$\forall x\in\Omega_t$ and $\forall \alpha\in\mathcal{S}$.

It may be concluded from the above analysis that there exists at
least a qualitative difference between the elastic domain
associated with rigidly rotating slip systems,
eq.(\ref{elastic1}), and the elastic domain associated with
convecting slip systems, eq.(\ref{elastic2}). More generally, each
type of slip systems evolution leads to a specific definition of
the yield surface. In all cases, the macroscopic yield surface may
be represented as the level set of a yield function having as
principal argument the rotated stress
$\bb{T}^{R}:=\left(\bb{R}^T\bb{T}\bb{R}\right)(t+\theta)$. The yield
surface associated with convecting slip systems is special;
we shall distinguish it by adding the subscript $_L$; thus we
write
\begin{equation}
f^R_{L}(\bb{T}^{R},...) = 0 \,\,\,\,\,\,\makebox{and}\,\,\,\,\,\, f^R_{W}(\bb{T}^{R},...) = 0
\end{equation}
for the macroscopic yield surfaces generated by assuming convective and, respectively, any other type of slip systems evolution at constituent level. Dots are to represent any additional macro-variables that may be required for the description of the yield surface,
e.g. hardening variables, structural tensors, etc. The superscript $^R$ is meant to indicate that the stress argument of the function
is the rotated stress;
by defining $f_W(\bb{T},...):=f^R_W(\bb{T}^R,...)$, and $f_L(\bb{T},...):=f^R_L(\bb{T}^R,...)$,
one obtains corresponding descriptions of the yield surface in terms of the Kirchhoff stress (referred to the laboratory frame).
Obviously, $f_W$ and $f_L$ must be isotropic functions; in case of anisotropic plastic properties,
the case of most interest in practice, this implies the presence of additional arguments in the form of structural
tensors, e.g., \cite{Boehler}, \cite{Liu}; in this sense, the Schmid stress may be regarded as the prototype example
(the dot product of two tensors being invariant to orthogonal transformations).

\subsection{The relationship between the macroscopic rate of plastic deformation and the yield surface. Plastic spin}

\noindent
No matter what characterization is obtained for the macroscopic yield surface, a generic yield surface described
as $f^R(\bb{T}^R,...) = 0 = f(\bb{T},...)$,
has the property that, for elastic unloading from a current yielding state $\bb{T}(t)$,
along an arbitrary stress path $\bb{T}(t+\theta)$, there holds $f^R(\bb{T}^R(t+\theta),...)- f^R(\bb{T}^R(t),...)<0$, for $\theta>0$,
whence, by dividing by $\theta$ and taking the limit $\theta\rightarrow 0$:
\begin{equation}
\hspace{-20pt}
\left.\frac{d}{d\theta}f^R(\bb{T}^R,...)\right|_{\theta=0}\leq 0
\iff \delta\bb{T}^R\cdot\frac{\partial f^R}{\partial\bb{T}^R}(\bb{T}^R(t))\leq0
\iff \delta^J\bb{T}\cdot\frac{\partial f}{\partial\bb{T}}(\bb{T}(t))\leq 0
\label{ygradient}
\end{equation}
Implicit in the above argument is that, besides the stress state, no other variable of the yield function is affected
during elastic unloading. It is assumed that the yield surface is smooth; thus
the gradient is well defined in the classical sense and at least part of the set of elastic directions $\delta\bb{T}^R = \bb{R}^T(t)\left(\delta^J\bb{T}\right)\bb{R}(t)$ spans the half space bounded by the tangent hyperplane at $\bb{T}(t)$.

Let again $\bb{T}(t)$ denote the current macro-stress state; it is assumed that $\bb{T}(t)$ is a yielding state.
Then given an arbitrary elastic direction $\delta\bb{T}^R$, originating at the current stress, by eq.(\ref{invarianceMacro}) there holds
\begin{equation}
\delta\bb{T}^R\cdot\left[\bb{R}^T(t)\bb{D}^p(t)\bb{R}(t)\right] = \frac{\overline{J}}{|\Omega_t|}\int_{\Omega_t}\frac{1}{J}\,\delta^L\bb{\tau}\cdot\bb{d}^{pL}\,dx
\end{equation}
In case of convecting slip systems, the integrand is always less than or equal to zero, by eq.(\ref{flowrule2}),
and hence, in the context of eq.(\ref{ygradient}), the macroscopic rate of deformation is along the exterior normal to the yield surface:
\begin{equation}
\delta\bb{T}^R\cdot\left(\bb{R}^T\bb{D}^p\bb{R}\right)\leq 0,\,\,\,\forall\,\delta\bb{T}^R\,\,\,\,\Longrightarrow\,\,\,\,
\bb{D}^p = \dot{\lambda}\,\bb{R} \frac{\partial f^R_L}{\partial\bb{T}^R}(\bb{T}^R)\bb{R}^T =
\dot{\lambda}\,\frac{\partial f_L}{\partial\bb{T}}(\bb{T})
\end{equation}
the scalar $\dot{\lambda}>0$ characterizing the magnitude of $\bb{D}^p$.

In the general case, of slip system evolution, let us employ the representation in eq.(\ref{psi}), for the Jaumann rate, in the inequality in eq.(\ref{flowrule}),
then substitute in the result the representation
$\delta^J\bb{\tau} = \bb{B}^J:\delta^J\bb{T}$, obtained from
eq.(\ref{stress1}), and then average over the current domain of the aggregate, to obtain
\begin{equation}
\delta^J\bb{T}\cdot\left\{\frac{\overline{J}}{|\Omega_t|}
\int_{\Omega_t}\frac{1}{J}\left(\bb{B}^J\right)^T:
\left[\bb{d}^p+\left(\bb{K}^J\right)^{-1}:\bb{\Psi}\right]\,dx\right\}\leq 0
\label{macrorigid}
\end{equation}
Let us define the symmetric second order tensor $\widehat{\bb{D}}^p$ by
\begin{equation}
\widehat{\bb{D}}^p:=\frac{\overline{J}}{|\Omega_t|}
\int_{\Omega_t}\frac{1}{J}\left(\bb{B}^J\right)^T:
\left[\bb{d}^p+\left(\bb{K}^J\right)^{-1}:\bb{\Psi}\right]\,dx
\label{jaumannMacro3}
\end{equation}
Inequality (\ref{macrorigid}) becomes\, $\delta^J\bb{T}\cdot\widehat{\bb{D}}^p\leq 0$;
hence the following normality rule holds:
\begin{equation}
\delta\bb{T}^R\cdot\left(\bb{R}^T\widehat{\bb{D}}^p\bb{R}\right)\leq 0,\,\,\,\forall\,\delta\bb{T}^R\,\,\,\,\Longrightarrow\,\,\,\,
\widehat{\bb{D}}^p = \dot{\lambda}\,\bb{R} \frac{\partial f^R_W}{\partial\bb{T}^R}(\bb{T}^R)\bb{R}^T =
\dot{\lambda}\,\frac{\partial f_W}{\partial\bb{T}}(\bb{T})
\label{rigidnormality}
\end{equation}

To interpret the above flow rule, let us first note that
employing the definition of the macro-rate $\bb{D}^p$ in eq.(\ref{jaumannMacro}),
the representation of $\bb{D}^{pL}$ in eq.(\ref{macrorateDP}), the definition of the tensor $\bb{B}^L$ in eq.(\ref{BTensor}),
and finally the relationship between $\bb{d}^p$ and $\bb{d}^{pL}$ in eq.(\ref{Truesdelldp}), with
the symmetry of the tensors $\bb{K}^J$, $\overline{\bb{K}}^J$ and $\bb{K}^L$, $\overline{\bb{K}}^L$,
one obtains the following representation of $\bb{D}^p$ in terms of microstructure events:
\begin{equation}
\bb{D}^p = \frac{\overline{J}}{|\Omega_t|}\int_{\Omega_t}\frac{1}{J}\left(\bb{B}^J\right)^T:\bb{d}^p\,dx
\label{jaumannMacro2}
\end{equation}
Next, let us define the symmetric second order tensor $\overline{\bb{\Psi}}$ by
\begin{equation}
\begin{array}{l}
\dd\overline{\bb{\Psi}}:=\overline{\bb{K}}^J:\left\{\frac{\overline{J}}{|\Omega_t|}
\int_{\Omega_t}\frac{1}{J}\left(\bb{B}^J\right)^T:
\left[\left(\bb{K}^J\right)^{-1}:\bb{\Psi}\right]\,dx\right\} =\vp\\
\dd = \frac{\overline{J}}{|\Omega_t|}
\int_{\Omega_t}\frac{1}{J}\left(\bb{A}^S\right)^T:\bb{\Psi}\,dx
\end{array}
\label{Psibar}
\end{equation}
where in the last equality the definition of the tensor $\bb{B}^J$ in eq.(\ref{stress1}) has been employed.
With eqs.(\ref{jaumannMacro3}), (\ref{jaumannMacro2}) and (\ref{Psibar}),
the macro-rate of plastic deformation admits the decomposition
$\bb{D}^p = \widehat{\bb{D}}^p - \left(\overline{\bb{K}}^J\right)^{-1}:\overline{\bb{\Psi}}$,
and eq.(\ref{rigidnormality}) becomes:
\begin{equation}
\bb{D}^p+\left(\overline{\bb{K}}^J\right)^{-1}:\overline{\bb{\Psi}} = \dot{\lambda}\,\frac{\partial f_W}{\partial\bb{T}}(\bb{T})
\label{flowmacro}
\end{equation}
This is the flow rule relating the direction of the rate of plastic deformation, $\bb{D}^p$,
with the exterior normal of the yield surface $f_W$.
Thus, besides the yield surface, an additional macro-variable, $\overline{\bb{\Psi}}$,
is required to completely characterize the direction of the rate of plastic deformation.

An alternative interpretation will lead us to the concept of plastic spin.
Let $V_T$ denote the subspace of symmetric second order tensors that commute with the current macro-stress $\bb{T}$,
that is, $V_T:=\left\{\bb{D}\,|\, \bb{T}\bb{D}=\bb{D}\bb{T}\right\}$;
since $V_T$ consists of all second order tensors that have the same eigenvectors as $\bb{T}$,
it is a three dimensional space.
Let $V_T^{\bb{\bot}}$ denote its orthogonal complement;
the macro-variable $\overline{\bb{\Psi}}$ admits the orthogonal decomposition $\overline{\bb{\Psi}} = \overline{\bb{\Psi}}_T+\overline{\bb{\Psi}}_T^{\bb{\bot}}$,
where $\overline{\bb{\Psi}}_T\in V_T$ and $\overline{\bb{\Psi}}_T^{\bb{\bot}}\in V_T^{\bb{\bot}}$.
One can then define an antisymmetric second order tensor $\bb{W}^p$ as the solution of the equation
\begin{equation}
\bb{T}\bb{W}^p-\bb{W}^p\bb{T} = \overline{\bb{\Psi}}_T^{\bb{\bot}}
\label{macroPlasticSpin}
\end{equation}
Substituting the decomposition $\bb{D}^p = \widehat{\bb{D}}^p - \left(\overline{\bb{K}}^J\right)^{-1}:\overline{\bb{\Psi}}$
and eq.(\ref{macroPlasticSpin}) into eq.(\ref{jaumannMacro}), leads to the following reformulation of the stress-strain relationship (\ref{jaumannMacro}) and of the flow rule (\ref{flowmacro}):
\begin{equation}
\bb{\dot{T}}^{Je} := \bb{\dot{T}}+\bb{T}\bb{W}^e-\bb{W}^e\bb{T}= \overline{\bb{K}}^J:\left(\bb{D}-\widehat{\widehat{\bb{D}}}^p\right)
,\,\,\,\,
\makebox{where}\,\,\,\,\bb{W}^e:=\bb{W}-\bb{W}^p
\label{macroflowrule1}
\end{equation}
\begin{equation}
\widehat{\widehat{\bb{D}}}^p +\left(\overline{\bb{K}}^J\right)^{-1}:\overline{\bb{\Psi}}_T = \dot{\lambda}\,\frac{\partial f_W}{\partial\bb{T}}(\bb{T})
\label{macroflowrule2}
\end{equation}
where $\widehat{\widehat{\bb{D}}}^p:=\widehat{\bb{D}}^p-\left(\overline{\bb{K}}^J\right)^{-1}:\overline{\bb{\Psi}}_T^{\bb{\bot}}$.
The tensor $\bb{W}^p$ is what is usually referred to as the plastic spin, e.g., \cite{Dafalias98}.
There is, however, a qualitative difference between the above defined plastic spin
and the one mentioned in the literature; the latter is always assumed,
either based on analogies inspired by the physics of plastic deformation at crystal level, e.g., \cite{Mandel82} or \cite{Dafalias90},
or by phenomenological axiomatization, e.g., \cite{Dafalias2000},
to describe an overall rigid body rotation of the yield surface, with respect to the material,
while the rate of plastic deformation is collinear with the exterior normal to the yield surface.
This is equivalent with stating that $\overline{\bb{\Psi}} = \overline{\bb{\Psi}}_T^{\bb{\bot}}$, or that $\overline{\bb{\Psi}}_T = \bb{0}$,
and hence with $\overline{\bb{\Psi}}\cdot\bb{D} = 0$, for any $\bb{D}\in V_T$;
letting $\bb{d}^D = \bb{A}^S:\bb{D}$ denote the elastic field generated via BVP(\ref{BVPelastic2}) by the application
of $\bb{D}$ on the boundary of the aggregate, and employing the definition in eq.(\ref{Psibar}) and the representation
in eq.(\ref{psi}), this is further equivalent to
\begin{equation}
0 = \int_{\Omega_t}\frac{1}{J}\,\bb{\Psi}\cdot\bb{d}^D\,dx = \int_{\Omega_t}\frac{1}{J}\,\psi\,dx,\,\,\,\forall \bb{D}\in V_T
\label{psiaverage}
\end{equation}
Thus, a plastic spin in the sense defined currently in the literature exists if and only if the
deviation from normality averages to zero for all fields $\bb{d}^D$. This is unlikely to happen for arbitrary
deformations.

To see this, let us consider two particular cases.
For rigidly rotating slip systems, with eq.(\ref{psirigid}), the above equality becomes:
\begin{displaymath}
0 = \int_{\Omega_t}\left(\bb{\sigma}\bb{w}^p-\bb{w}^p\bb{\sigma}\right)\cdot\bb{d}^D\,dx =
\int_{\Omega_t}\left(\bb{\sigma}\bb{d}^D-\bb{d}^D\bb{\sigma}\right)\cdot\bb{w}^p\,dx,\,\,\,\forall \bb{D}\in V_T
\end{displaymath}
Given the heterogeneity of $\bb{w}^p$, for an arbitrary polycrystal,
it is likely that the condition $\bb{\sigma}\bb{d}^D=\bb{d}^D\bb{\sigma}$, for all $\bb{D}$,
is not only sufficient, but also necessary;
considering, for the purpose of illustration, that $(1/J)\bb{K}^L$ is isotropic and constant,
there holds $\bb{d}^D=\bb{D}$; then given the heterogeneity of the current stress field, it is likely
that $\bb{\sigma}\bb{D}\neq\bb{D}\bb{\sigma}$ holds for at least some non-negligible subset of $\Omega_t$.

For unit convecting slip systems, recalling that $\bb{G}_A = -\bb{I}\otimes\bb{m}^{\alpha}\otimes\bb{m}^{\alpha}$
and $\bb{G}_B = \bb{I}\otimes\bb{n}^{\alpha}\otimes\bb{n}^{\alpha}$,
substituting the representation of $\bb{\Psi}$ in eq.(\ref{psi}) into eq.(\ref{psiaverage}) obtains:
\begin{displaymath}
0 = \int_{\Omega_t}\frac{1}{J}\,\sum_{\alpha\in\mathcal{A}}\dot{\gamma}^{\alpha}\tau^{\alpha}_{cr}
\left[\left(\bb{d}^D:\bb{n}^{\alpha}\right)\cdot\bb{n}^{\alpha} - \left(\bb{d}^D:\bb{m}^{\alpha}\right)\cdot\bb{m}^{\alpha}\right]\,dx,\,\,\,\forall \bb{D}\in V_T
\end{displaymath}
Again, given the heterogeneity of the plastic deformation at constituent level, it is likely that
the above condition holds for arbitrary deformations only if the bracketed object vanishes (almost) everywhere;
this is possible only if $\bb{d}^D \equiv \bb{I}$.

Hence the plastic spin, as defined by eq.(\ref{macroPlasticSpin}) is always accompanied by a deviation
of the macro-rate of plastic deformation from the gradient of the yield surface.
When structural changes (in the form of texture effects) are neglected, that is, when it assumed that $\bb{w}^p\equiv\bb{0}$,
one has $\bb{\Psi}\equiv\bb{0}$ in the case of rigidly rotating slip systems; hence the plastic spin and the deviation
from the gradient of the yield surface both vanish, and the macro-model recovers the normality structure.
This is not the case for all types of slip systems evolution;
for unit convecting slip systems, neglecting texture effects does not imply that $\bb{\Psi} \equiv \bb{0}$,
and hence, with or without texture effects considered, the structure of the macro-model remains the same.


\section{Conclusions}

\noindent
Assuming slip is the only mechanism of plastic deformation at crystal level and that Schmid law governs slip activity,
the basic structure of the rate equations modeling the plastic
deformation of metals at macroscopic level can be deduced, by entirely rigorous arguments,
and under general conditions, from the properties of the constituent crystals.
Among these, the evolution of the slip systems of a constituent has
been found to have a significant effect upon the structure of the macroscopic constitutive model.
In any case, a macroscopic rate of deformation $\bb{D}^p$ can be defined unequivocally in terms of
constituent properties and this is related to the Jaumann stress rate $\bb{\dot{T}}^J:=\bb{\dot{T}}+\bb{T}\bb{W}-\bb{W}\bb{T}$
of the Kirchhoff macro-stress $\bb{T}$ by the classical relationship
\begin{equation}
\bb{\dot{T}}^J = \bb{K}^J:\left(\bb{D}-\bb{D}^p\right)
\label{Z1}
\end{equation}
where $\bb{K}^J$ is the corresponding tensor of instantaneous elastic response, symmetric and positive definite,
and $\bb{D}$ and $\bb{W}$ are the macro-rate of deformation and spin.

In an Eulerian framework, the macroscopic yield surface can be represented as $f(\bb{T},...) = 0$,
with dots representing the rest of the arguments of the yield function $f$, which
may be required for a complete and accurate model (hardening variables, structural tensors, etc).
In general, the yield surface depends on the type of slip system evolution
adopted at crystal level and, with one exception, the direction of the macro-rate $\bb{D}^p$ features deviations from the exterior normal
to the yield surface at the current stress-state.

Only if the slip systems convect with the lattice, the corresponding yield surface, say $f_L(\bb{T},...)=0$,
is such that the classical normality rule holds:
\begin{equation}
\bb{D}^p = \dot{\lambda}\,\frac{\partial f_L}{\partial \bb{T}}
\end{equation}
with $\dot{\lambda}$ a scalar multiplier characterizing the magnitude of $\bb{D}^p$ (via the consistency condition).

For any other type of slip system evolution, there exists an additional
macro-variable, denoted here $\overline{\bb{\Psi}}$, such that the following relationship between
the macro-rate of plastic deformation and the exterior normal to the yield surface, described as $f_W(\bb{T},...) = 0$,
holds:
\begin{equation}
\bb{D}^p+\left(\bb{K}^J\right)^{-1}:\overline{\bb{\Psi}} = \dot{\lambda}\,\frac{\partial f_W}{\partial \bb{T}}
\label{Z2}
\end{equation}

An alternative formulation can be obtained if one considers the
decomposition $\overline{\bb{\Psi}} = \overline{\bb{\Psi}}_T+\overline{\bb{\Psi}}_T^{\bb{\bot}}$,
where $\overline{\bb{\Psi}}_T^{\bb{\bot}}$ is the component of $\overline{\bb{\Psi}}$ that is
orthogonal to the subspace of symmetric second order tensors that commute with the current macro-stress $\bb{T}$.
Then there exists $\bb{W}^p$, an antisymmetric tensor, solution of the equation
\begin{equation}
\bb{T}\bb{W}^p-\bb{W}^p\bb{T} = \overline{\bb{\Psi}}_T^{\bb{\bot}}
\end{equation}
and, by defining $\widetilde{\bb{D}}^p=\bb{D}^p-\left(\bb{K}^J\right)^{-1}:\overline{\bb{\Psi}}_T$, and $\bb{W}^e:=\bb{W}-\bb{W}^p$,
the constitutive system (\ref{Z1}), (\ref{Z2}) acquires the equivalent reformulation
\begin{equation}
\bb{\dot{T}}^{Je}:= \bb{\dot{T}}+\bb{T}\bb{W}^e-\bb{W}^e\bb{T} = \bb{K}^J:\left(\bb{D}-\widetilde{\bb{D}}^p\right)
\label{Z1b}
\end{equation}
\begin{equation}
\widetilde{\bb{D}}^p+\left(\bb{K}^J\right)^{-1}:\overline{\bb{\Psi}}_T = \dot{\lambda}\,\frac{\partial f_W}{\partial \bb{T}}
\end{equation}
The tensor $\bb{W}^p$ is the plastic spin; contrary to what is commonly assumed,
it is always accompanied by a deviation of the macro-rate of plastic deformation from the normal to the yield surface,
that is, for general deformations there holds $\overline{\bb{\Psi}}_T\neq \bb{0}$.

\newpage

\appendix
\section{}
\textbf{Symmetry and positive definiteness of the tensor of instantaneous moduli}\\\\

\noindent In the context of Section \ref{SectionMacroElastic}, let
$\bb{L}^{\delta}_1$ and $\bb{L}^{\delta}_2$ denote two elastic
macro-velocity gradients, with corresponding local elastic fields
$\bb{l}^{\delta}_1$, $\delta^L_1\bb{\tau}$ and
$\bb{l}^{\delta}_2$, $\delta^L_2\bb{\tau}$, respectively. Applying
Hill's Lemma to the equilibrated field
$(1/J)\left(\delta^L_1\bb{\tau}+\bb{l}^{\delta}_1\bb{\tau}\right)$
and to the compatible field $\bb{l}^{\delta}_2$ results in
\begin{equation}
\left(\delta^L_1\bb{T}+\bb{L}^{\delta}_1\bb{T}\right)\cdot\bb{L}^{\delta}_2
= \frac{\overline{J}}{|\Omega_t|}
\int_{\Omega_t}\frac{1}{J}\left(\delta^L_1\bb{\tau}+\bb{l}^{\delta}_1\bb{\tau}\right)\cdot\bb{l}^{\delta}_2\,dx
\label{KLM2}
\end{equation}
By the (major) symmetry of $\bb{K}^L$ and the identity
$\left(\bb{l}^{\delta}_1\bb{\tau}\right)\cdot\bb{l}^{\delta}_2 =
\left(\bb{l}^{\delta}_2\bb{\tau}\right)\cdot\bb{l}^{\delta}_1$,
there holds
\begin{equation}
\hspace{-1cm}
\left(\delta^L_1\bb{\tau}+\bb{l}^{\delta}_1\bb{\tau}\right)\cdot\bb{l}^{\delta}_2
=
\left(\bb{K}^L:\bb{d}^{\delta}_1+\bb{l}^{\delta}_1\bb{\tau}\right)\cdot\bb{l}^{\delta}_2
=
\left(\bb{K}^L:\bb{d}^{\delta}_2+\bb{l}^{\delta}_2\bb{\tau}\right)\cdot\bb{l}^{\delta}_1
=
\left(\delta^L_2\bb{\tau}+\bb{l}^{\delta}_2\bb{\tau}\right)\cdot\bb{l}^{\delta}_1
\end{equation}
allowing us to write, by employing once again Hill's Lemma,
\begin{equation}
\hspace{-1cm}
\frac{\overline{\bb{J}}}{|\Omega_t|}\int_{\Omega_t}
\frac{1}{J}\left(\delta^L_1\bb{\tau}+\bb{l}^{\delta}_1\bb{\tau}\right)\cdot\bb{l}^{\delta}_2\,dx
= \frac{\overline{J}}{|\Omega_t|}\int_{\Omega_t}
\frac{1}{J}\left(\delta^L_2\bb{\tau}+\bb{l}^{\delta}_2\bb{\tau}\right)\cdot\bb{l}^{\delta}_1\,dx
=
\left(\delta^L_2\bb{T}+\bb{L}^{\delta}_2\bb{T}\right)\cdot\bb{L}^{\delta}_1
\end{equation}
Substituting into eq.(\ref{KLM2}), employing the identity
$\left(\bb{L}^{\delta}_1\bb{T}\right)\cdot\bb{L}^{\delta}_2 =
\left(\bb{L}^{\delta}_2\bb{T}\right)\cdot\bb{L}^{\delta}_1$ and
eq.(\ref{KLMacro}), results in
\begin{equation}
\left(\overline{\bb{K}}^L:\bb{D}^{\delta}_1\right)\cdot\bb{D}^{\delta}_2
=
\left(\overline{\bb{K}}^L:\bb{D}^{\delta}_2\right)\cdot\bb{D}^{\delta}_1
\end{equation}
identity valid for any symmetric second order tensors
$\bb{D}^{\delta}_i=(\bb{L}^{\delta}_i+\bb{L}^{\delta T}_i)/2$,
$i=1,2$, hence proving the (major) symmetry of
$\overline{\bb{K}}^L$.

To investigate the conditions under which $\overline{\bb{K}}^L$
may be positive definite, let us apply Hill's Lemma to the
equilibrated field
$(1/J)\left(\delta^L\bb{\tau}+\bb{l}^{\delta}\bb{\tau}\right)$ and
to the compatible field $\bb{l}^{\delta}$, both corresponding to
the boundary condition $\bb{L}^{\delta}=\bb{D}^{\delta}$, where
$\bb{D}^{\delta}$ is an arbitrary second order symmetric tensor\,:
$\left(\delta^L\bb{T}+\bb{D}^{\delta}\bb{T}\right)\cdot\bb{D}^{\delta}
=
(\overline{J}/|\Omega_t|)\int_{\Omega_t}(1/J)\left(\delta^L\bb{\tau}+
\bb{l}^{\delta}\bb{\tau}\right)\cdot\bb{l}^{\delta}\,dx$\,; since
$\delta^L\bb{T}\cdot\bb{D}^{\delta} =
\left(\overline{\bb{K}}^L:\bb{D}^{\delta}\right)\cdot\bb{D}^{\delta}$,
this identity may be rewritten in the form
\begin{equation}
\hspace{-1cm}
\left(\overline{\bb{K}}^L:\bb{D}^{\delta}\right)\cdot\bb{D}^{\delta}
=
\frac{\overline{J}}{|\Omega_t|}\int_{\Omega_t}\frac{1}{J}\left(\bb{K}^L:\bb{d}^{\delta}\right)\cdot\bb{d}^{\delta}\,dx
+\frac{\overline{J}}{|\Omega_t|}\int_{\Omega}\bb{\sigma}\cdot\left(\bb{l}^{\delta
T}\bb{l}^{\delta}\right)dx -
\bb{T}\cdot\left(\bb{D}^{\delta}\right)^2 \label{BarKpositive1}
\end{equation}
With $\bb{K}^L$ being positive definite, for
$\gamma:=\min\left\{\left(\bb{K}^L:\bb{d}\right)\cdot\bb{d},\,\,
||\bb{d}||=1\right\}>0$ we obtain by employing
eq.(\ref{ASlocalization}):
\begin{equation}
\frac{\overline{J}}{|\Omega_t|}\int_{\Omega_t}
\left(\bb{K}^L:\bb{d}^{\delta}\right)\cdot\bb{d}^{\delta}\,\frac{dx}{J}\geq
\frac{\overline{J}}{|\Omega_t|}\int_{\Omega_t}\gamma\bb{d}^{\delta}\cdot\bb{d}^{\delta}\,\frac{dx}{J}
=
\left(\overline{A}^S:\bb{D}^{\delta}\right)\cdot\bb{D}^{\delta}
\end{equation}
where
$\overline{A}^S:=\left(\overline{J}/|\Omega_t|\right)\int_{\Omega_t}\gamma\left(\bb{A}^S\right)^T\bb{A}^S
(dx/J)$; the tensor $\overline{A}^S$ is (symmetric and) positive
definite, since
$\left[\left(\bb{A}^S\right)^T\bb{A}^S:\bb{D}^{\delta}\right]\cdot\bb{D}^{\delta}
= ||\bb{d}^{\delta}||^2\geq0$. Then with
\begin{displaymath}
\overline{\gamma}:=\min\left\{\left(\overline{\bb{A}}^S:\bb{D}\right)\cdot\bb{D},\,\,
||\bb{D}||=1\right\}>0 
\end{displaymath}
there holds
\begin{equation}
\frac{\overline{J}}{|\Omega_t|}\int_{\Omega_t}
\frac{1}{J}\left(\bb{K}^L:\bb{d}^{\delta}\right)\cdot\bb{d}^{\delta}\,dx\geq\overline{\gamma}\,||\bb{D}^{\delta}||^2
\label{BarKpositive2}
\end{equation}
On the other hand, since
$\bb{l}^{\delta}(x)=\overline{A}(x):\bb{D}^{\delta}$, one can
write $\bb{\sigma}\cdot\left(\bb{l}^{\delta
T}\bb{l}^{\delta}\right) =
\left(\bb{N}:\bb{D}^{\delta}\right)\cdot\bb{D}^{\delta}$, where
$N_{abcd}:=\sigma_{ij}\overline{A}_{kiab}\overline{A}_{kjcd}$\,;
then
$(\overline{J}/|\Omega_t|)\int_{\Omega_t}\bb{\sigma}\cdot\left(\bb{l}^{\delta
T}\bb{l}^{\delta}\right)dx =
\left(\overline{\bb{N}}:\bb{D}^{\delta}\right)\cdot\bb{D}^{\delta}$,
where
$\overline{\bb{N}}:=(\overline{J}/|\Omega_t|)\int_{\Omega_t}\bb{N}\,dx$.
Since $\bb{D}\longrightarrow
\left(\overline{\bb{N}}:\bb{D}\right)\cdot\bb{D}$ is continuous,
there exists and is finite
$\eta:=\max\left\{\left|\left(\overline{\bb{N}}:\bb{D}\right)\cdot\bb{D}\right|,\,\,
||\bb{D}||=1\right\}$, and $\eta\geq 0$. Thus:
\begin{equation}
\left|\frac{\overline{J}}{|\Omega_t|}\int_{\Omega_t}\bb{\sigma}
\cdot\left(\bb{l}^{\delta
T}\bb{l}^{\delta}\right)dx\right|\leq\eta||\bb{D}^{\delta}||^2
\label{BarKpositive3}
\end{equation}
Next, by working with respect to a frame in which $\bb{T}$ is
diagonal, we have
\begin{equation}
\hspace{-21pt}
\bb{T}\cdot\left(\bb{D}^{\delta}\right)^2 =
\sum_{i}T_{ii}\left(D^{\delta}_{ii}\right)^2 =
\sum_{i}T_{ii}\sum_k\left(D^{\delta}_{ik}\right)^2\Longrightarrow
\left|\bb{T}\cdot\left(\bb{D}^{\delta}\right)^2\right|\leq
\mu\,||\bb{D}^{\delta}||^2,
\label{BarKpositive4}
\end{equation}
where $\mu:=\max\left\{\left|T_{ii}\right|,\, i=1,2,3\right\}$.
From eqs.(\ref{BarKpositive1}), (\ref{BarKpositive2}),
(\ref{BarKpositive3}) and (\ref{BarKpositive4}) it follows that a
sufficient condition for $\overline{\bb{K}}^L$ to be positive
definite is that: $\overline{\gamma}>(\eta+\mu)$. Let us observe
now that $\eta$ and $\mu$ are of the order of the current yielding
stress while $\overline{\gamma}$ is of the same order as the
elastic moduli. Since in metals the elastic moduli are two to
three orders of magnitude greater than the flow stresses, we can
safely assume that the above inequality is always satisfied and
hence conclude that $\overline{\bb{K}}^L$ is positive
definite.\\\\


%

\begin{thebibliography}{50}




\bibitem[Asaro(1983)]{Asaro}
Asaro, R.J., 1983. Crystal plasticity. J. Applied Mech., 50, 921-934.


\bibitem[Asaro and Rice(1977)]{AsaroRice}
Asaro, R.J., Rice, J.R., 1977. Strain localization in ductile single crystals. J. Mech. Phys. Solids, 25, 309-338.







\bibitem[Boehler(1987)]{Boehler}
Boehler, J.P., 1987. Applications of tensor functions in solid mechanics. CISM Courses and Lectures, vol. 292. Springer, Berlin.



\bibitem[Bunge and Nielsen(1997)]{Bunge97}
Bunge, H.J., Nielsen, I., 1997. Experimental determination of plastic spin in polycrystalline materials. Int. J. Plast., 13, 435-446.







\bibitem[Dafalias and Aifantis(1990)]{Dafalias90}
Dafalias, Y.F., Aifantis, E.C., 1990. On the microscopic origin of the plastic spin. Acta Mech., 82, 31--48.


\bibitem[Dafalias(1998)]{Dafalias98}
Dafalias, Y.F., 1998. Plastic spin: necessity or redundancy ? Int. J. Plast., 14, 909-931.


\bibitem[Dafalias(2000)]{Dafalias2000}
Dafalias, Y.F., 2000. Orientational evolution of plastic orthotropy in sheet metals.
J. Mech. Phys. Solids, 48, 2231-2255.









\bibitem[Gurtin et al(2009)]{GurtinAnand}
Gurtin, M., Fried, E., Anand, L., 2009. The mechanics and thermodynamics of continua. Cambridge University Press. \S 20.





\bibitem[Hill(1967)]{Hill67}
Hill, R., 1967.
The essential structure of constitutive laws for metal composites and polycrystals. J. Mech. Phys. Solids, 15, 79-95.

\bibitem[Hill(1968a)]{Hill68a}
Hill, R., 1968a.
On constitutive inequalities for simple materials-I. J. Mech. Phys. Solids, 16, 229-242.

\bibitem[Hill(1968b)]{Hill68b}
Hill, R., 1968b.
On constitutive inequalities for simple materials-II. J. Mech. Phys. Solids, 16, 315-322.

\bibitem[Hill(1972)]{Hill72}
Hill, R., 1972.
On constitutive macro-variables for heterogeneous solids at finite strain. Proc. R. Soc. Lond. A., 326, 131-147.



\bibitem[Hill and Rice(1972)]{HillRice72}
Hill, R., Rice, J.R., 1972.
Constitutive analysis of elastic-plastic crystals at arbitrary strain. J. Mech. Phys. Solids, 20, 401-413.

\bibitem[Hill and Rice(1973)]{HillRice73}
Hill, R., Rice, J.R., 1973.
Elastic potentials and the structure of inelastic constitutive laws. SIAM J. Appl. Math., 25, 448-461.




\bibitem[Kim and Yin(1997)]{KimYin}
Kim, K.H., Yin, J.J., 1997. Evolution of anisotropy under plane stress. J. Mech. Phys. Solids, 45, 841-851.


\bibitem[Liu(1982)]{Liu}
Liu, I.-Shih., 1982. On representations of anisotropic invariants. Int. J. Eng. Sci., 20, 1099-1109.


\bibitem[Mandel(1982)]{Mandel82}
Mandel, J., 1982. Definition d'un repere privilegie pour l'etude des transformations anelastique du polycristal.
J. Mec. Theo. Appl., 1, 7-23.

\bibitem[Mehrabadi and Cowin(1990)]{MCowin}
Mehrabadi, M.M., Cowinn, S., 1990. Eigentensors of linear anisotropic materials. Q. J. Mech. Appl. Math., 43, 15-41.

\bibitem[Prantil et al(1993)]{Prantil93}
Prantil, V.C., Jenkins, J.T., Dawson, P.R., 1993.
An analysis of texture and plastic spin for planar polycrystals. J. Mech. Phys. Solids, 41, 1357-1382.









\bibitem[Schmid(1924)]{Schmid1924}
Schmid, E., 1924. Zn - normal stress law. In Proc. Intern. Congr. Appl. Mech. Delft, 342-252.












\bibitem[Spitzig and Richmond(1984)]{SR84}
Spitzig, W.A., Richmond, O., 1984. The effect of pressure on the flow stress of metals.
Acta Metall., 32, 457-463.




\bibitem[Taylor(1938)]{Taylor38}
Taylor, G.I., 1938. Plastic strain in metals. J. Inst. Metals, 62, 307-324.

\bibitem[Tugcu and Neale(1999)]{Tugcu}
Tugcu, P., Neale, K.W., 1999. On the implementation of anisotropic yield functions into finite strain problems
of sheet metal forming. Int. J. Plast., 15, 1021-1040.



















\end{thebibliography}



\end{document}